\documentclass[aps,prl,twocolumn,superscriptaddress]{revtex4-1}

\usepackage{amsmath,amssymb,color,graphicx,mathpazo,mathtools,times,float}
\usepackage[colorlinks=true,allcolors=blue]{hyperref}
\usepackage{soul}
\usepackage{upgreek,cleveref}
\usepackage{subcaption}

\newcommand{\comment}[1]{}

\newcommand{\eref}[1]{Eq.~\eqref{eq:#1}}

\newcommand{\Fref}[1]{Figure~\ref{fig:#1}}
\newcommand{\fref}[1]{Fig.~\ref{fig:#1}}

\begin{document}
\title{Floquet control of optomechanical bistability in multimode systems}

\author{Karl Pelka}
\email{karl.a.pelka@um.edu.mt}
\affiliation{Department of Physics, University of Malta, Msida MSD 2080, Malta}
\author{Guilhem Madiot}
\email{guilhem.madiot@icn2.cat}
\affiliation{Centre de Nanosciences et de Nanotechnologies, CNRS, Universit\'e Paris-Saclay, Palaiseau, France}
\author{R\'emy Braive}
\affiliation{Centre de Nanosciences et de Nanotechnologies, CNRS, Universit\'e Paris-Saclay, Palaiseau, France}
\affiliation{Universit\'e de Paris, F-75006 Paris, France}
\author{Andr\'e Xuereb}
\affiliation{Department of Physics, University of Malta, Msida MSD 2080, Malta}

\date{\today}

\begin{abstract}

Cavity optomechanical systems enable fine manipulation of nanomechanical degrees of freedom with light, adding operational functionality and impacting their appeal in photonic technologies. We show that distinct mechanical modes can be exploited with a temporally modulated laser drive to steer between bistable steady states induced by changes of cavity radiation pressure. We investigate the influence of thermo-optic nonlinearity on these Floquet dynamics and find that it can inhibit or enhance the performance of the coupling mechanism in contrast to their often performance limiting character. Our results provide new techniques for the characterization of thermal properties and the control of optomechanical systems in sensing and computational applications.

\end{abstract}

\maketitle

\emph{Introduction.}---Cavity optomechanics employs optical forces to exert control over optical and mechanical degrees in micro-mechanical systems and is a contemporary research field with outstanding progress~\cite{Aspelmeyer2014}. Prototypically, an optomechanical system consists of a single mode of the electromagnetic radiation field, e.g., within a high-finesse optical cavity~\cite{Kippenberg2005}, interacting with the motion of a harmonic oscillator by means of the radiation pressure force~\cite{Marquardt2006}. The optomechanical interaction has been used to cool the motion of the mechanical system down to its ground state~\cite{Chan2011,Teufel2011} and generate quantum entanglement between mechanical oscillators~\cite{Ockeloen2018,Riedinger2018}. On the other hand, it is also possible to transfer energy from the optical field into the mechanical oscillator which leads to self-sustained oscillations and lies at the heart of synchronization phenomena in optomechanics~\cite{Heinrich2011,Lauter2017,Lipson2012,Holmes2012,Amitai2017,Loerch2017,Lauter2015,Lipson2015,Colombano2019,Pelka2020,Madiot2020}. \par
Such systems may also find technological use; synchronized optomechanical arrays, for example, could act as high-power and low-noise on-chip frequency sources~\cite{Lipson2015}, proof-of-concept isolators and directional amplifiers for microwave radiation were produced~\cite{Bernier2017,Malz2018,Barzanjeh2017,Mercier2019} as well as bidirectional conversion between microwave and optical light was shown~\cite{Andrews2014}. Other potential application of uniformly driven optomechanical systems lie in the non-linear behaviour which can create an effective double-well potential for the mechanical degree of freedom resulting in the optomechanical bistability~\cite{Dorsel1983,Ghobadi2011}. Nanomechanical elements which can controllably be put into distinct mechanical states can act as memory cells which are quintessential for possible nanomechanical computing devices~\cite{Badzey2004,Maboob2008,Bagheri2011}. As these devices reach the nanoscale this can cause competitive information densities which can be operated at frequencies in the GHz range~\cite{Badzey2004}. In addition, an optomechanical realization will be operated fully optically while being resistant to magnetic perturbations~\cite{Bagheri2011}.\par
The study of non-uniform optical driving schemes resulted recent advances in optomechanics driven by theoretical advances with the Floquet approach~\cite{Malz2016,Pietikainen2020}. It enables non-reciprocal transfer of phonons~\cite{Xu2019} leading to topological transport of phonons via synthetic gauge fields~\cite{Peano2015,Walter2016,Mathew2018}, allows quantum states to be transferred from one mechanical element to another \cite{Weaver2017}, and to entangle such elements \cite{Ockeloen2018}. These temporal control schemes also allow to overcome mode-competition inhibiting multiple mechanical modes to simultaneously experience amplification resulting in mode-locked lasing of degenerate modes~\cite{Mercade2021}. Additionally, recent studies investigated the characterization of the cavity's thermal properties based on Floquet techniques and measurement effects on the quantum mechanical properties in the mechanical ground state~\cite{Eichenfield2009,Verhagen2012,Li2014,Qiu2019,Ma2021}. In this letter, we show that the Floquet driving approach offers dynamical control of the optomechanical bistability in multimode settings which presents a useful tool in the manipulation of optomechanical systems. We derive a spectral method that incorporates thermo-optical effects which suggest that thermo-optical effects can inhibit or even improve the control and underpin its predictions with experimental results. Finally, we explore their use for elementary phononic memory elements, frequency sensing and discuss logic elements generalizations.

\emph{Model.}---We consider the collective dynamics of a system consisting of $N$ mechanical modes coupled to one optical mode, which is described by the optomechanical Hamiltonian
\begin{equation}
\label{eq:OMModel}
\hat{H}_{\text{S}}/\hbar=\omega_{\text{op}}\hat{a}^{\dagger}\hat{a}+\sum\limits_{j=1}^N(\Omega_j\hat{b}_j^{\dagger}\hat{b}_j-g_{j}\hat{a}^{\dagger}\hat{a}(\hat{b}_j+\hat{b}_j^{\dagger})).
\end{equation}
with $\hat{a}$ ($\hat{b}_j$) the optical (mechanical) annihilation operator, $\omega_\text{op}$ ($\Omega_j$) the corresponding resonance frequencies, and $g_j$ the vacuum optomechanical coupling rates. The laser driving the optics is modelled by extending the Hamiltonian with $i\hbar[\mathcal{E}_{\text{drive}}(t)\hat{a}^{\dagger}-\mathcal{E}_{\text{drive}}^*(t)\hat{a}]$, where the driving laser $\mathcal{E}_{\text{in}}=\mathcal{E}_{0}e^{i\omega_{\text{L}}t}$ is subject to optical modulation $\mathcal{E}_{\text{drive}}(t)=\mathcal{E}_{\text{in}}(t)\mathcal{T}(t)$. We assume a Mach--Zehnder modulator (MZM) whose transfer characteristic $\mathcal{T}(t)=e^{i\phi_{0}}(1+e^{i\phi_{\text{mod}}(t)})/2$ implements intensity modulation for $\phi_{\text{mod}}(t)=-\pi/2+d\cos(\theta(t))$ and can be expressed in terms of the Bessel functions of the first kind $\mathcal{J}_n(d)$ using the Jacobi-Anger expansion
\begin{equation}
\frac{\mathcal{T}(t)}{e^{i\phi_0}}=\frac{1-i\mathcal{J}_0(d)}{2}+\sum_{n=1}^{\infty}i^{n+1}\mathcal{J}_n(d)\cos(n\theta(t)),
\end{equation}
with $\theta(t)=\Omega_{\text{mod}}t+\theta_0$. This indicates that an increasing modulation depth $d$ involves increasingly many driving tones beyond the usual first order expansion~\cite{Eichenfield2009,Verhagen2012,Li2014,Qiu2019,Ma2021,Allain2021}. \par
Employing the standard procedure of appending bath degrees of freedom and tracing them out~\cite{Aspelmeyer2014} results in quantum Langevin equations. These are separable into mean field and fluctuation components  ($\hat{a}(t)e^{i\omega_{\text{L}}t+i\phi_0}=\alpha(t)+\hat{\mathfrak{a}}(t)$ and $\hat{b}_j(t)=\beta_j(t)+\hat{\mathfrak{b}}_j(t)$), with mean fields obeying
\begin{align}
\dot{\alpha}&=\biggl\{-i\biggl[\Delta-\sum\limits_{j=1}^Ng_j\text{R}(\beta_j)\biggr]-\frac{\kappa}{2}\biggr\}\alpha+\mathcal{E}_0\mathcal{T}e^{-i\phi_0}, \nonumber \\
\dot{\beta}_j&=-\bigg(i\Omega_{j}+\frac{\Gamma_j}{2}\bigg)\beta_j+ig_j|\alpha|^2.
\label{eq:MeanFieldEvolution}
\end{align}
Here, $\Delta=\omega_{\text{op}}-\omega_{\text{L}}$ denotes the detuning of the central laser frequency from the optical resonance and $\text{R}(z)=z+z^*$.

In addition to the dispersive optomechanical coupling, the cavity in experimental setups absorbs photons and heats up which in turn changes its refractive index and geometry. We acknowledge and model the heating process by the dynamics of the temperature deviation $\delta \dot{T}(t)=g_{\text{abs}}|\alpha|^2(t)-\gamma_{\text{th}}\delta T (t)/2$ and the resulting shift of the optical cavity frequency $\omega_{\text{op}}\approx\omega_{\text{op}}(\bar{T})+\frac{\partial\omega_{\text{op}}}{\partial T}(T(t)-\bar{T}) =\omega_0+g_{\text{T}}\delta T(t)$, due to this photo-thermo-refractive-shift mechanism (PTRS)~\cite{Eichenfield2009,Verhagen2012,Li2014,Qiu2019,Ma2021,Allain2021}. Here, $g_{\text{abs}}$ denotes the temperature change due to linear photon absorption, $\gamma_{\text{th}}$ the thermalization rate, and $g_T$ parametrizes the linear thermal shift of the optical frequency.
The mean field equation for the mechanical field $\beta(t)$ and the temperature deviation $\delta T(t)$ can be solved in terms of the mean intensity $|\tilde{\alpha}|^2(\omega)$ in Fourier space.
Since the equation for the mean optical field $\alpha$ is periodic in time, we choose a Floquet ansatz and express $\alpha$ as a truncated Fourier series $\alpha(t)=\sum_n\alpha_ne^{-in\Omega_{\text{mod}}t}$ with $n\in\{-D,...,D\}$. The resulting intensity of the mean field is then $|\tilde{\alpha}|^2(\omega)=\sum_{(p,q)}\alpha_p\alpha^*_{p-q}\delta(\omega-iq\Omega_{\text{mod}})$, where $p\in\{-D,...,D\}$ and $q\in\{-D+p,...,D+p\}$.
The Floquet ansatz results in the dynamical system
\begin{equation}
\dot{\alpha}_m=\mathcal{E}_0\mathcal{T}_m-\chi^{-1}_{m}\alpha_{m} +\sum_{(p,q)}\chi^{-1}_{\text{cub},q}\alpha_{p}\alpha_{p-q}^*\alpha_{m-q},
\end{equation}
such that acquiring the steady state ($\dot{\alpha}_m=0$) amounts to solving $4D+2$ coupled real cubic equations. 
Here, we defined $\mathcal{T}_0=(1-i\mathcal{J}_0(d))/2$, $\mathcal{T}_m=i^{|m|+1}\mathcal{J}_m(d)$, $\chi^{-1}_{m}=i(\Delta-m\Omega_{\text{mod}})+\kappa/2$, and $\chi^{-1}_{\text{cub},q}=\chi^{-1}_{\text{Th},q}+\sum_{j}\chi^{-1}_{\text{OM},jq}$ with $2\pi \chi^{-1}_{\text{Th},q}/g_{\text{T}}g_{\text{abs}}=(q\Omega_{\text{mod}}-i\gamma_{\text{th}}/2)^{-1}$ and $2\pi \chi^{-1}_{\text{OM},jq}/g_j^2=[i(q\Omega_{\text{mod}}-\Omega_j)-\Gamma_j/2]^{-1} -[i(q\Omega_{\text{mod}}+\Omega_j)+\Gamma_j/2]^{-1}$.
Techniques from algebraic geometry~\cite{Hassett2007} yield an analytic result for $D=0$ and numerical methods are necessary to obtain the steady state $\bar{\alpha}_m$ beyond $D=0$. \par
The resulting $\bar{\alpha}_m$ turn the dynamics of fluctuation components $\hat{\mathfrak{a}}$ and $\hat{\mathfrak{b}}$ into a periodic system which can be treated with Floquet techniques~\cite{Malz2016} up to leading order
\begin{align}
\dot{\hat{\mathfrak{a}}}^{(0)}=&\tilde{\chi}^{-1}\hat{\mathfrak{a}}^{(0)}+\sum\limits_{n}\sum\limits_{j=1}^N ig_j\bar{\alpha}_{n}\mathfrak{R}(\hat{\mathfrak{b}}_j^{(n)})+\sqrt{\kappa}\hat{\mathfrak{a}}_{\text{in}}^{(0)}, \nonumber \\
\dot{\hat{\mathfrak{b}}}_j^{(n)}=&\check{\chi}_{jn}^{-1}\hat{\mathfrak{b}}^{(n)}_j+ig_j\bigl[\bar{\alpha}_{-n}^*\hat{\mathfrak{a}}^{(0)}+\bar{\alpha}_n\hat{\mathfrak{a}}^{\dagger(0)}\bigr]+\sqrt{\Gamma_j}\hat{\mathfrak{b}}_{j,\text{in}}^{(n)}, 
\label{eq:FloqDyn}
\end{align}
with the mechanical Floquet susceptibilities $\check{\chi}_{jn}^{-1}=-[i(\Omega_j-n\Omega_{\text{mod}})+\Gamma_j/2]$ and the optical susceptibility $\tilde{\chi}^{-1}=-[i(\Delta-\sum_j g^2_j|\bar{\alpha}_0|^2I(\check{\chi}_{j0}^{-1})/|\check{\chi}_{j0}^{-1}|^2)+\kappa/2]$ where we denote $\mathfrak{R}(\hat{o})=\hat{o}+\hat{o}^{\dagger}$ and $I(z)=i(z^{*}-z)$.
 Using the input--output relations for the relevant contributions to the optical field $\hat{\mathfrak{a}}_{\text{out}}(\omega)=\hat{\mathfrak{a}}^{(0)}_{\text{in}}(\omega)-\sqrt{\kappa}\hat{\mathfrak{a}}^{(0)}(\omega)$ with input noise obeying $\langle\hat{\mathfrak{u}}^{(m)}_{\text{in}}(\omega)\hat{\mathfrak{w}}^{\dagger(p)}_{\text{in}}(\omega')\rangle=\delta(\omega-\omega')\delta_{\mathfrak{u}\mathfrak{w}}\delta_{mp}(\mathfrak{n}_{\text{th}}^{\mathfrak{u}}+1)$ yields the stationary power spectral density of the experimentally accessible output field 
\begin{equation}
\label{eq:6}
S(\omega)=\tilde{S}+\sum\limits_{n,j}\frac{\kappa g^2_j|\bar{\alpha}_{n}|^2\Gamma_j\bar{n}_j}{\Big[(\omega-\bar{\Delta})^2+\frac{\kappa}{4}\Big]\Big[(\omega-\Omega_{jn})^2+\frac{\Gamma_j^2}{4}\Big]}
\end{equation}
consisting of a noise floor $\tilde{S}$ and multiple Lorentzian peaks at $\Omega_{jn}=\Omega_j+n\Omega_{\text{mod}}$. In sideband unresolved systems ($\kappa \gg \Omega_{jn}$), these are filtered equally by the Lorentzian cavity density of states with effective detuning $\bar{\Delta}=\Delta+\sum_{j,n}2g_j^2|\bar{\alpha}_n|^2/\Omega_j$, due to static radiation pressure for $\Gamma_j \ll \Omega_j$. Consequently, the spectrum displays the mean field amplitudes $|\bar{\alpha}_n|^2$ in leading order.

\begin{figure}
\includegraphics[width=0.45\textwidth,angle=0]{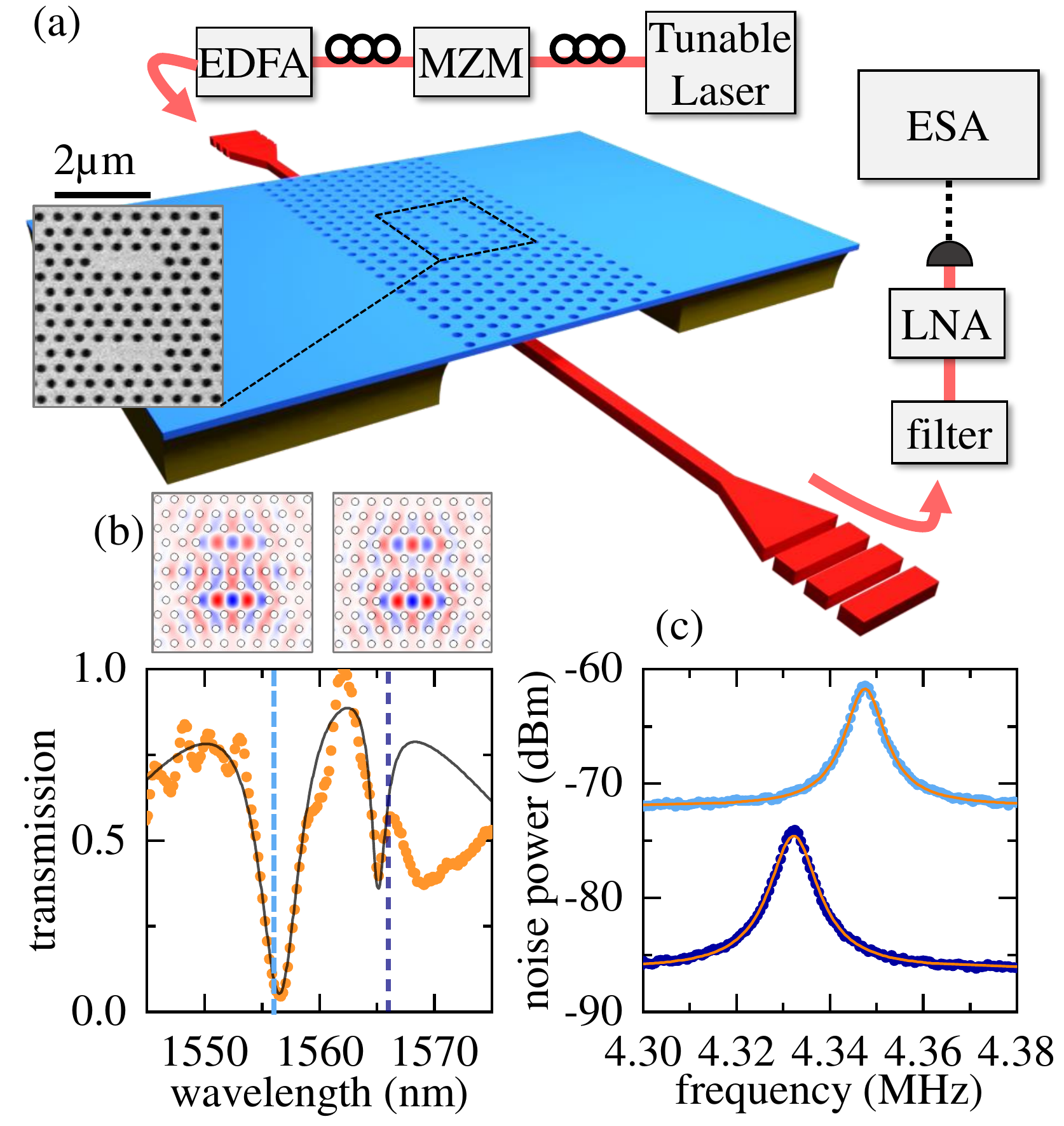}
\caption{(Color online) (a) Experimental setup including the integrated optomechanical platform with a suspended 2D photonic crystal InP nanomembrane (blue) above a SOI waveguide (red). inset: SEM micrograph of the photonic crystal molecule. (b) Measured transmission spectrum (orange dots) an fit (black line). The electric field distribution associated to each photonic resonance are shown on top. (c) Mechanical spectrum measured by optomechanically probing either the photonic mode (-) (dark blue) or the mode (+) (light blue).}
\label{fig:fig1}
\end{figure}

\begin{figure*}
\includegraphics[width=0.999\textwidth,angle=0]{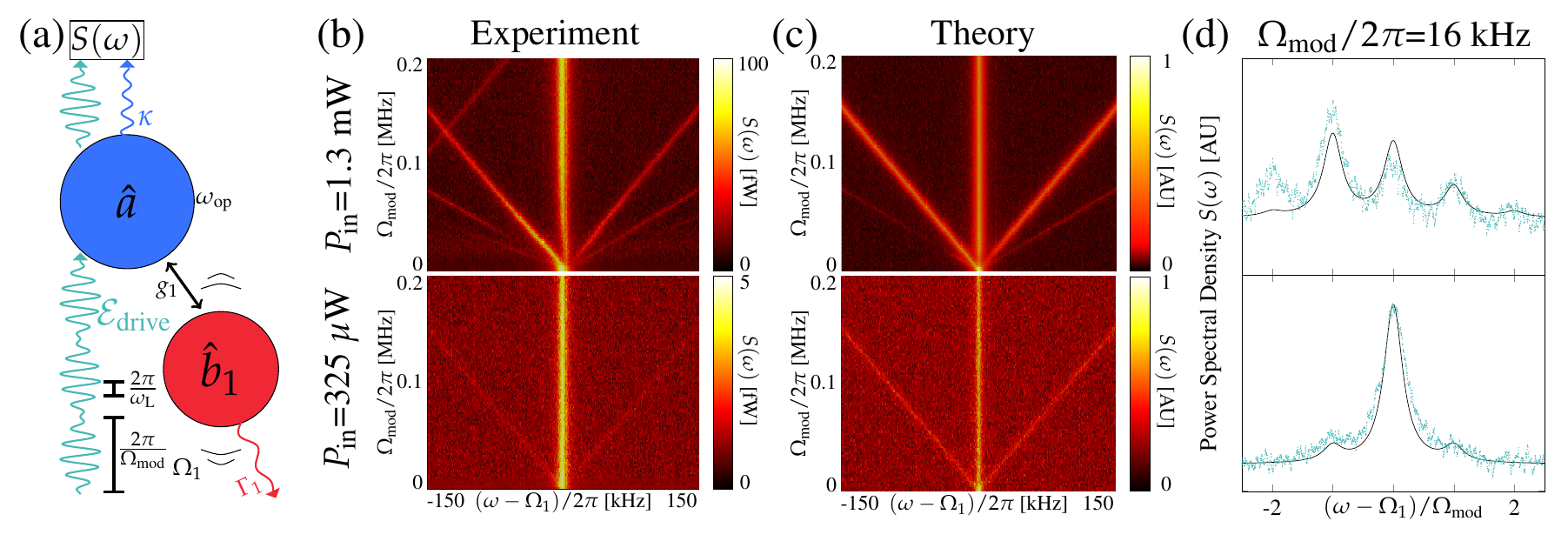}
\caption{(Color online) Floquet dynamics of a single-mode optomechanical system (a) Schematic of an optomechanical cavity driven with a modulated laser field. (b) Experimentally measured noise spectra centered at the mechanical frequency $\Omega_1/2\pi=4.330$ MHz for $P_{in}=1.3$ mW (top) and $P_{in}=325$ $\mu$W (bottom) mapped over the modulation frequency $\Omega_{\text{mod}}$. (c) Theoretically predicted noise spectra for varying $\Omega_{\text{mod}}$ employing high (top) and low (bottom) optical power. (d) Horizontal section of the respective diagrams of measured spectra (green) and numerical results (black) at $\Omega_\text{mod}/2\pi=16$ kHz.}
\label{fig:Phasespace}
\end{figure*}
\emph{Results with one mechanical mode.}---Aiming to observe the model dynamics with one mechanical mode, we use a 265 nm thin InP $10\times20$ $\upmu$m$^2$ membrane suspended over a rib silicon waveguide via a 250 nm air-gap illustrated in \fref{fig1} (a). 
The membrane is pierced with a 2D photonic crystal at the center of which two L3 defect cavivites are designed. These defects, shown in the inset of \fref{fig1} (a), allow localized photonic modes to be evanescently driven from the waveguide. 
The optical channel transmission spectrum is measured by injecting a broadband light source into the waveguide gratings termination. The transmitted field is collected and sent to a monochromator. The normalized transmission spectrum is plotted in \fref{fig1} (b). We fit the data using the coupled mode theory (CMT) of two waveguide-coupled photonic cavities~\cite{Li2010} and ignore the right most feature. From the fit, the bonding and antibonding modes central wavelengths are found to be respectively $\lambda_- = 1557.27$ nm and $\lambda_+ = 1565.55$ nm, with total quality factors $Q_-^{tot} \approx 380$ and $Q_+^{tot} \approx 3240$. The discrepancy between fit and data around 1570 nm is due to imperfect alignment of the injection and collection fiber tips with regard to the SOI gratings. The distributions of the electric field transverse component are simulated for both modes and shown in \fref{fig1} (b). We place the chip in a vacuum chamber pumped below 10$^{-5}$ mbar and perform all the following measurements at room temperature.

To access the mechanical noise spectrum of the suspended membrane, a tunable laser resonantly drives a given optical mode (dashed vertical lines in \fref{fig1} (b)). The output signal is filtered, sent to a low noise amplifier (LNA), and coupled to a low-sensitivity photodetector. We measure the resulting RF signal with an electrical spectrum analyzer (ESA). The suspended membrane sustains several mechanical modes with frequencies ranging from 4 MHz to more than 100 MHz. These resonances are coupled with the optical modes through optomechanical couplings of dissipative and dispersive nature~\cite{Tsvirkun2015}. As illustrated in \fref{fig1} (c), the mechanical spectrum can be accessed by driving either the bonding (light blue) or antibonding (dark blue) optical mode. In this work we focus on the fundamental mode with central frequency $\Omega_1/2\pi = 4.330$ MHz and mechanical linewidth $\Gamma_1/2\pi = 6$ kHz.

Before injecting into the system, the laser with wavelength $\lambda_L=1565.75$ nm passes a Mach--Zehnder modulator (MZM) in which we input a RF signal $V(t) = V_\mathrm{mod}\cos\Omega_\mathrm{mod}t$. The modulation depth is $d=\pi\times V_\mathrm{mod}/V_\pi$ with the calibrated half-wave voltage $V_\pi=7.0$ V. We record the output optical field noise spectrum as illustrated in \fref{Phasespace} (a). The resulting experimental diagrams using a modulation depth of $d=0.89$ are depicted in \fref{Phasespace} (b). The top figure shows the result for the input power $P_{\text{in}}=1.3$ mW which corresponds to the center of the previously characterized thermo-optic bistability (see Supplemental Material). We observe modulation sidebands surrounding the mechanical peak, with imbalanced amplitudes due to thermo-optical effects. For comparison, the identical measurement realized in the low-power situation is shown in the bottom of \fref{Phasespace} (b). In this case, only one pair of sidebands with weak and balanced amplitudes are recorded. The numerical prediction by \eref{6} with $g_1=1$ MHz, $\Omega_1/g_1=4.34$, $\Gamma_1/g_1=6 \times 10^{-3}$, $\kappa/g_1=5.45 \times 10^{3}$, $\Delta/g_1=6.60 \times 10^{3}$ neglecting higher order contributions (See Supplemental Material) is presented in the top of \fref{Phasespace} (c) showing qualitative agreement with the experiment at large input power. We employ a drive of $\mathcal{E}_0/g_1=1.85\times 10^{4}$ and modulation depth $d=1.35$ in addition with the thermo-optical coupling strength $g_{\text{T}}g_{\text{abs}}/g^2_1=2.3$ and thermalization rate $\gamma_{\text{th}}=4.25 \mu$s. We find that $\Omega_\mathrm{mod}$ allows control over the transduced modulation comb. This effect requires sufficiently high input power and modulation frequencies below 125 kHz. This cut-off frequency finds its origins in the thermalization rate of the material. In an independent measurement (see Supplemental Material), we measure the switching transition time of approximately $4$ $\mu$s in the thermo-optic resonator, in good agreement with previous measurements in a similar device
~\cite{Brunstein2009}. Higher modulation frequency suppresses the thermo-optic effect. Consequently, the modulation comb retains its symmetry. We perform a measurement as a function of the modulation depth (see Supplemental Material) and find that this parameter also enables control over the modulation comb asymmetry. Numerical simulations of \eref{6} with a reduced driving strength $\mathcal{E}_0/g_1=10^{3}$ and modulation depth $d=0.89$ shown in the bottom of \fref{Phasespace} (c) agree with the experimental result and show only one pair of symmetric sidebands. Horizontal sections for a fixed modulation frequency of $\Omega_{\text{mod}}=16$ kHz of the respective theoretical (black) and experimental (green) heatmaps are shown in \fref{Phasespace} (d) for large (top) and low (bottom) input power further confirm the observed model dynamics.

\begin{figure}[t]
\begin{center}
\includegraphics[width=0.475\textwidth,angle=0]{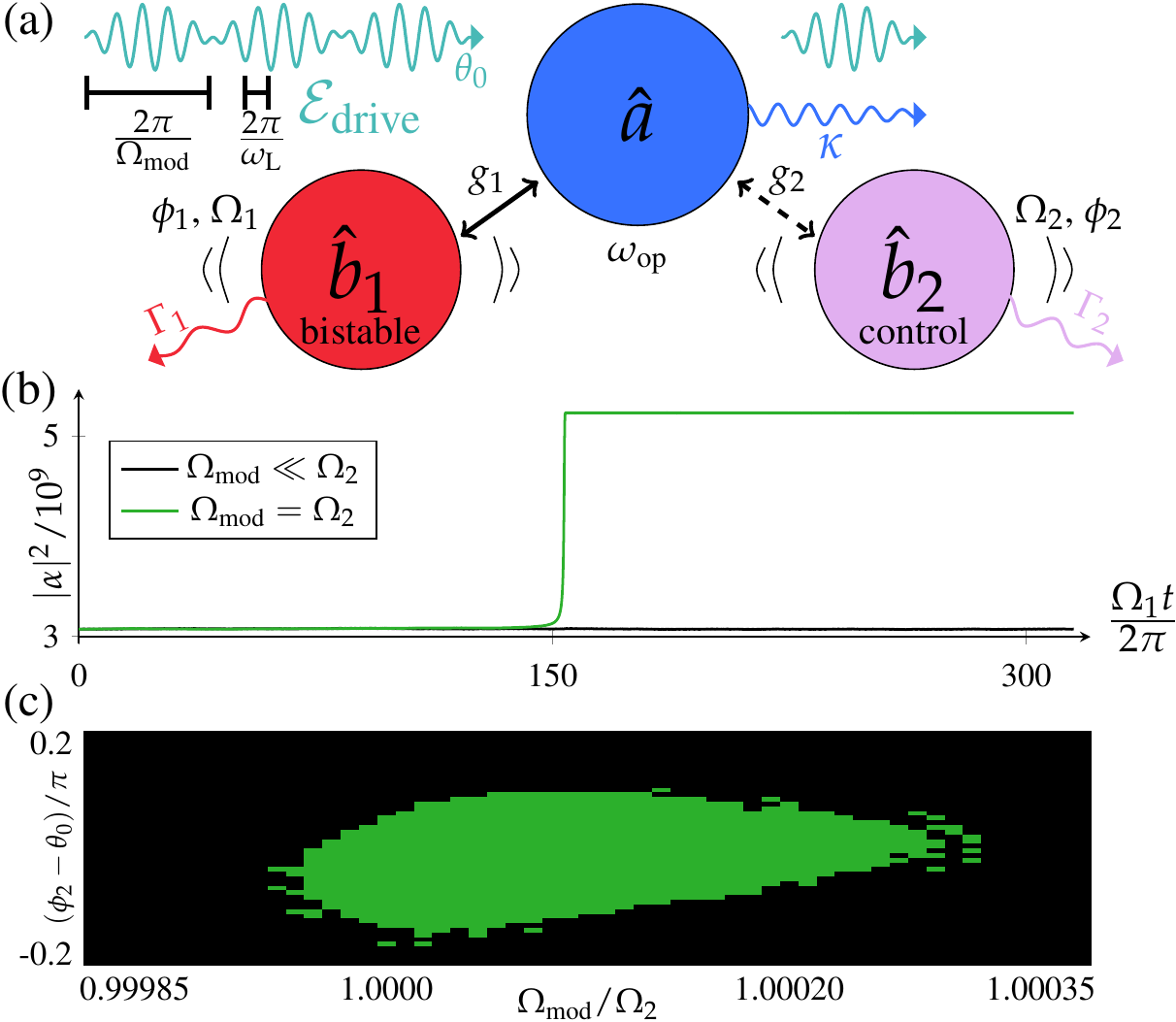}
\end{center}
\caption{(Color online) Floquet control of optomechanical bistability. (a) An optomechanically bistable mode is controlled by a second, degenerate control mode with an intensity-modulated pump. (b) The steady state switches when the modulation frequency $\Omega_{\text{mod}}$ matches the control frequency $\Omega_2$ (green) whereas it is unaffected for off-resonant modulation (black). (c) Two-dimensional parameter diagram describing the switching capability of the Floquet drive: The intensity modulation's phase $\theta_0$ aligning with the mechanical phase $\phi_2$ in addition to the frequency matching leads to steady state switching (green).}
\label{fig:Numerics1}
\end{figure}

\emph{Floquet control of optomechanical bistability.}---Based on our model and its agreement with experiment for one mechanical mode, we extend the discussion to potential applications with multimode systems.
We can analyze the interaction of the mechanical Floquet modes $\hat{\mathfrak{b}}_j^{(m)}$ mediated through the optical field fluctuations by eliminating $\hat{\mathfrak{a}}^{(0)}$ and find their effective coupling via the contributions
\begin{equation}
\sigma_{jlp}^{(m)}(\omega)=\frac{g_jg_l\bar{\alpha}^{*}_{-m}\bar{\alpha}_{p}}{i(\bar{\Delta}-\omega)+\frac{\kappa}{2}}-\frac{g_jg_l\bar{\alpha}_{m}\bar{\alpha}^{*}_{p}}{-i(\bar{\Delta}+\omega)+\frac{\kappa}{2}}.
\end{equation}
The stationary mechanical spectra without periodic drive ($m\equiv p \equiv 0$) are Lorentzians~\cite{Genes2009,Karuza2012} $S_{\hat{\mathfrak{b}}_j}(\omega)=\tilde{S}_{\hat{\mathfrak{b}}_j}+\Gamma_j\bar{n}_j[(\Omega'_j-\omega)^2+\Gamma'^2_j/4]^{-1}$ with optical-spring-corrected frequencies $\Omega'_j=\Omega_j\sqrt{\textrm{I}(\sigma_{jj0}^{(0)}(\Omega_j))/4+1}$ and modified linewidths $\Gamma'_j=\Gamma_j+\textrm{R}(\sigma_{jj0}^{(0)}(\Omega_j))$. 
The former expression allows assessing the stability of mechanical oscillators' steady states for red-detuned driving ($\bar{\Delta}>0$): If we examine the static frequency response we find $\Omega'_j(\omega=0)=\Omega_j\sqrt{\eta_j}$ with $\eta_j=1-\bar{\Delta} g_j^2|\bar{\alpha}_0|^2/[\Omega_j(\bar{\Delta}^2+\kappa^2/4)]$ which has to be larger than zero for a stable steady state in accordance with the standard treatment via the Routh-Hurwitz criterion~\cite{Genes2009}. In the presence of the periodic drive, there are additional contributions to the frequency response which modify the stability parameter
\begin{equation}
\tilde{\eta}_j=\eta_j+\frac{1}{4}\sum\limits_{n,l \ne j}\text{I}\bigg[\frac{\sigma_{jln}^{(0)}\sigma_{jl0}^{(n)}}{i(\Omega_l-n\Omega_{\text{mod}})+\frac{\Gamma_l}{2}+\sigma_{jln}^{(n)}}\bigg].
\label{eq:control}
\end{equation}
This suggests that a mechanical mode $\hat{b}_l$ can influence the occurence of the optomechanical bistability of a distinct mechanical mode $\hat{b}_j$ if the modulation frequency is tuned into resonance at $\Omega_{\text{mod}}=\Omega_l/n$ with $n\in \mathbb{Z}$ on the scale of the mechanical linewidth $\Gamma_l$.
We investigate the predicted capability of the periodic drive in \eref{control} to control the bistability of a distinct mechanical mode as depicted in \fref{Numerics1} (a). We therefore conduct numerical simulations with system parameters which exhibit an optomechanical bistability based on~\cite{Ghobadi2011}. It consists of a mechanical oscillator with frequency $\Omega_1=10$ MHz, damping rate $\Gamma_1=500$ kHz, and mass $m_1=5$ ng, coupled to a Fabry--P\'{e}rot cavity of length $L=1$ mm and finesse $\mathcal{F}=1.07 \times 10^{4}$ with the strength $g_1=\omega_c/L\sqrt{\hbar/m\Omega_1}$, driven by a laser with $\lambda=810$ mm and $\Delta=2.62\Omega_1$ and $\mathcal{E}_0/\sqrt{2}=6.4\times 10^{6}$. Additionally, a second mechanical mode with frequency $\Omega_2=11$ MHz, damping rate $\Gamma_2=55$ kHz and coupling strength $g_2/g_1=5.53 \times 10^{-2}$ is used to control the prior one's steady state. We inspect the effect of the modulated drive with modulation depth $d=1.875 \times 10^{-5}$ to the mean field dynamics of the It\^{o} stochastic differential equation corresponding to \eref{MeanFieldEvolution}. We study thermal excitation corresponding to shot noise $n^{\mathfrak{a}}_{\text{th}}=n^{\mathfrak{b}_1}_{\text{th}}=0$ for the cavity and the bistable mechanical mode and $n^{\mathfrak{b}_2}_{\text{th}}=8000$ phonons with examples depicted in \fref{Numerics1} (b) employing the Euler--Maruyama scheme~\cite{Kloeden1992}. The system remains stable in its steady state for off-resonant modulation $\Omega_{\text{mod}}=1$ MHz $\ll \Omega_2$. For sufficient time under resonant modulation $\Omega_{\text{mod}}=\Omega_2$, switching of the steady state occurs (see Supplemental Material) and enables the setup to detect and signal the frequency $\Omega_2$ in the signal fed into the MZM. \Fref{Numerics1} (c) summarizes the result of omitting thermal excitation and replacing it with periodic drive to clarify the switching mechanism: Switching of the steady state occurs if the phase $\phi_2$ of the mechanical oscillator used to control the bistability aligns with the phase $\theta_0$ of the optical modulation for resonant intensity modulation. This requires the control oscillator to assume the correct phase for sufficiently long optical modulation (See Supplemental Material) which is caused by phase noise and shows the necessity of thermal excitation. Tuning the modulation depth, we find that the amplitude of the sidebands $\bar{\alpha}_{n}$ can be increased or suppressed for modulation frequencies in the thermo-optical regime (See Supplemental Material). Since \eref{control} suggests that the underlying coupling strength grows (non-linearly) with these amplitudes, \emph{photothermal effects and thermal excitation can be exploited for increased control} of multimode optomechanical systems.

\emph{Conclusions.}---Our investigation reveals that thermal properties of optomechanical systems can be employed to tailor its Floquet dynamics. Using a 2D sideband unresolved optomechanical photonic crystal, we demonstrated experimentally how a Kerr-type nonlinearity---namely the thermo-optic effect---can achieve the predicted desymmetrization. This method conveniently characterizes thermal properties which we verify with  independent measurements. Interestingly such nonlinearities are ubiquitous in semiconductor microcavities, with cut-off frequencies ranging from a few kHz and surpassing the GHz range \cite{Pelc:14}, depending on the process nature. These Floquet modes allow to control the bistability of a distinct mechanical mode which can be understood from higher-order cross-mode contributions to the self-energy with modulated drive. The mechanism is shown with two mechanical modes where the thermal excitation of one mode allows resonant modulation to trigger a response of the other. This mechanism applies equally to multiple harmonically spaced control modes where the switching can implement logical rules.

\section*{Acknowledgments}
This work is supported by the European Union's Horizon 2020 research and innovation program under Grant Agreement No.\ 732894 (FET Proactive HOT), the French RENATECH network, the Agence Nationale de la Recherche as part of the “Investissements d’Avenir” program (Labex NanoSaclay, ANR-10-LABX-0035) with the flagship project CONDOR and the JCJC project ADOR (ANR-19-CE24-0011-01).
\appendix

\section{Corrections to the power spectral density of higher-order Floquet modes}
\label{sec:AppA}
The experimentally recorded spectra show additional imbalance of the modulation sidebands which cannot be explained in terms of the leading order description. Therefore, we inspect the linearized fluctuation dynamics
\begin{align}
\dot{\hat{\mathfrak{a}}}&=-\bigg(i\Delta+\frac{\kappa}{2}\bigg)\hat{\mathfrak{a}}-i\sum\limits_{j=1}^Ng_j(\alpha\mathfrak{R}(\hat{\mathfrak{b}}_j)+\hat{\mathfrak{a}}\text{R}(\beta_j))+\sqrt{\kappa}\hat{\mathfrak{a}}_{\text{in}}, \nonumber \\
\dot{\hat{\mathfrak{b}}}_j&=-\bigg(i\Omega_{j}+\frac{\Gamma_j}{2}\bigg)\hat{\mathfrak{b}}_j-ig_j(\alpha^*\hat{\mathfrak{a}}+\alpha\hat{\mathfrak{a}}^{\dagger})+\sqrt{\Gamma_j}\hat{\mathfrak{b}}_{j,\text{in}}.
\end{align}
The periodic mean field $\alpha(t)=\sum_n\bar{\alpha}_ne^{-in\Omega_{\text{mod}}t}$ allows to expand the fluctuation dynamics in terms of Floquet modes
\begin{align}
\dot{\hat{\mathfrak{a}}}^{(m)}=&-\chi_{m}^{-1}\hat{\mathfrak{a}}^{(m)}-\sum\limits_{(p,q)}\sum\limits_{j=1}^N \chi_{\text{OM},jq}^{-1}\bar{\alpha}_p\bar{\alpha}^{*}_{p-q}\hat{\mathfrak{a}}^{(m-q)} \nonumber \\
&-\sum\limits_{n=-D}^D\sum\limits_{j=1}^N ig_j\bar{\alpha}_{-n}\mathfrak{R}(\hat{\mathfrak{b}}_j^{(m-n)})+\sqrt{\kappa}\hat{\mathfrak{a}}_{\text{in}}^{(m)}, \nonumber \\
\dot{\hat{\mathfrak{b}}}_j^{(m)}=&-\tilde{\chi}_{\text{me},m}^{-1}\hat{\mathfrak{b}}^{(m)}_j-ig_j\sum_{n=-D}^{D}(\bar{\alpha}_{-n}^*\hat{\mathfrak{a}}^{(m-n)}+\bar{\alpha}_n\hat{\mathfrak{a}}^{\dagger(m-n)})\nonumber \\
&+\sqrt{\Gamma_j}\hat{\mathfrak{b}}_{j,\text{in}}^{(m)},
\end{align}
with $\tilde{\chi}_{\text{me},m}^{-1}=i(\Omega_j-m\Omega_{\text{mod}})+\Gamma_j/2$. Restricting to $\hat{\mathfrak{a}}^{(0)}$ results in Eq. (5) in the main text. Including the higher order fluctuation modes results in the Fourier transform
\begin{align}
\hat{\mathfrak{a}}^{(0)}(\omega)=\frac{\sqrt{\kappa}\hat{\mathfrak{a}}^{(0)}_{\text{in}}(\omega)-\sum\limits_{p=-D}^D\sum\limits_{j=1}^N\frac{ig_j\bar{\alpha}_{p}\sqrt{\Gamma_j}\hat{\mathfrak{b}}_{j,\text{in}}^{(p)}(\omega)}{\chi^{-1}_{\text{me},-p}-i\omega}}{\chi_{0,\text{cav}}^{-1}-i\omega+\sum\limits_{p=-D}^{D}\sum\limits_{j=1}^N|\bar{\alpha}_p|^2\chi^{-1}_{\text{OS},pj}(\omega)}
\end{align}
which shows that the optomechanical interaction alters the optical detuning and decay rate by $\chi^{-1}_{\text{OS},pj}=\chi^{-1}_{\text{OM},j0}+\chi^{-1}_{\text{OM},j}(\omega+p\Omega_{\text{mod}})$ where the former contribution is frequency independent and leads to the static optical spring effect covered in the main text. The latter contributions however make the effective detuning $\tilde{\Delta}(\omega)=\bar{\Delta}+\sum_{j,p}|\bar{\alpha}_p|^2\text{Im}(\chi^{-1}_{\text{OM},j}(\omega+p\Omega_{\text{mod}}))$ and decay $\tilde{\kappa}(\omega)=\kappa+\sum_{j,p}2|\bar{\alpha}_p|^2\text{Re}(\chi^{-1}_{\text{OM},j}(\omega+p\Omega_{\text{mod}}))$ frequency dependent which will also be reflected in the accessible power spectral density of the output field
\begin{align}
S(\omega)=\tilde{S}+\sum\limits_{p,j}\frac{\kappa g^2_j|\bar{\alpha}_{p}|^2\Gamma_j\bar{n}_j}{\Big[(\omega-\tilde{\Delta})^2-\frac{\tilde{\kappa}^2}{4}\Big]\bigg[(\omega-\Omega_{jp})^2+\frac{\Gamma_j^2}{4}\bigg]}.
\end{align}
These effects modify the cavity density of states and lead to a change of the apparent imbalance of the mean field amplitudes $|\alpha_n|^2$ displayed by the power spectral density. These contributions were not included in the numerical analysis of the experiment as they made the numerical fitting procedure unstable.

\section{Thermo-optic effect and thermalization time}
\label{sec:AppB}

The physical origin of the thermo-optic effect in our experiment is the temperature growth in the material induced by light absorption which is responsible for a significant shift of the dielectric index. In an optical cavity, this effect is enhanced such that it can red-shift the cavity resonance frequency. If the input field intensity passes a certain threshold, the resonance lineshape becomes bistable. Such behavior can be evidenced by scanning forward and backward the laser frequency over the resonance, or equivalently, by sweeping up and down the input laser intensity.

We use a tunable laser and inject light into the waveguide through the aligned injection fibers. The output laser field is sent to a low-power photodetector and the DC response is checked on an oscilloscope. Therefore, the waveguide transmission is now triggered in real-time, provided that the transmission can be re-normalized. The input power is estimated by measuring the off-resonance transmission $\zeta\approx0.1$ of the integrated waveguide and assuming the injection and the collection efficiency to be equal. The input power is therefore $P_\mathrm{in}=\sqrt{\zeta}P_\mathrm{inj}$ with the optical power sent in the injection fiber $P_\mathrm{inj}$.

\begin{figure}
\begin{center}
\includegraphics[scale=0.9,trim=5cm 1.3cm 5cm 2cm]{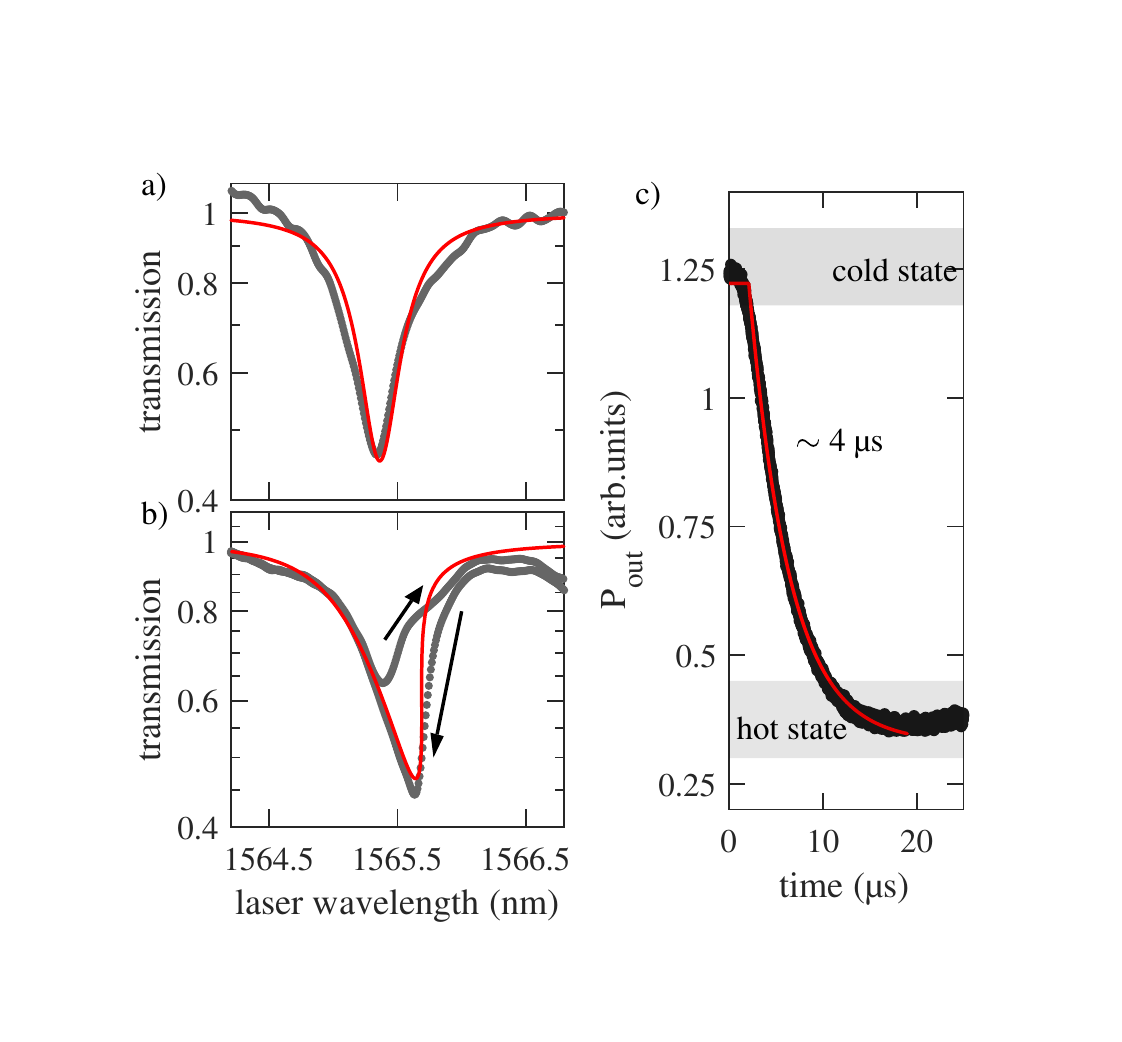}
\caption{a) Spectral transmission response using CW laser with input power $P_\mathrm{in}\approx325$ $\mu$W. The data (grey dots) are fitted with a CMT model (red line). b) Idem using $P_\mathrm{in}\approx1.3$ mW. Here both forward and backward scans evidence an hysteretic behaviour due to thermo optic nonlinearity. c) Averaged dynamical response of the photonic mode under 10 kHz square modulation of the input laser set to the bistability center. Measurements (black dots) are fitted with ring-down and built-in exponential functions (red) returning respectively a cooling time of 3.5 µs and a heating time of 4.4 µs.}
\label{fig:S1}
\end{center}
\end{figure}

For low power the observed transmission dip can be fitted with the linear transmission expression such that the internal and external Q-factors are determined. In \fref{S1} (a), with $P_\mathrm{in}=325$ $\mu$W, we find $Q_i\approx4400$ and $Q_\mathrm{w}\approx9500$.
The measurement is reproduced using both forward and backward scans of the laser wavelength at $P_\mathrm{in}\approx1.3$ mW. We fit the data with a nonlinear CMT model implementing a linear dependence of the resonance wavelength with the cavity temperature. 
Although the fit accurately matches with the width of the observed dip, and also retrieves the presence of a bistable region, we note a disagreement in the size of the bistability. We attribute this discrepancy to a too large scanning speed of the laser wavelength. In practice, it is set at 10 nm/s in order to prevent oscillations in the laser output power, which would have corrupted the measured transmission. This results in an averaging effect of the transmission near the bistability edges. In the experimental data, the jumps of the optical states are not abrupt as expected, but follow the photodetector response lifetime ($\approx6$ ms).

In the thermo-optic bistability, the optical resonator intra-cavity intensity is likely to switch stable state due to external perturbation such as e.g. noise or input field modulation. The switching time $\uptau_s$ is given by the thermalization time of the resonator.
Under sufficiently strong external modulation, the resonator can switch periodically, at the modulation frequency. However if the latter is higher than a certain cut-off frequency, given by $(2\uptau_s)^{-1}$, the resonator cannot switch twice a modulation period. This cut-off frequency therefore defines a limitation for the processes relying of thermo-optic nonlinearity.
In order to estimate the switching time $\uptau_s$, the input laser is modulated at sufficiently low-frequency for the transition regime to be observed. For this purpose, the laser wavelength is set at the center of the bistability ($\lambda = 1566.75$ nm) and modulated in the MZM with a square signal carrying amplitude $V_\mathrm{mod}=2$ V and frequency $\omega_\mathrm{mod}=10$ kHz. At the waveguide output, a fiber splitter allows to trigger the transmitted signal via a fW sensitive photodetector.

Using a modulation depth $d=0.89$ and frequency $\Omega_{\text{mod}}=10$ kHz, we record the optical output and average hundreds of modulation periods. The data are shown in \fref{S1}(c). Here, the optical resonator intra-cavity field switches from the cold state (high transmission) to the hot state and then returns back to the cold state at half a cycle following an exponential decay. We fit the data with a function $f(t)=A\exp(-t/\uptau_s) + B$ which provides the thermalization time $\uptau_s\approx4$ $\mu$s. Following the above discussion, we deduce that the corresponding cut-off frequency is of the order of 125 kHz.

\section{Numerical simulation procedures demonstrating bistability control}
\label{sec:AppD}
The numerical procedure that we use to generate the sample trajectories of our model displayed in Fig. 3 (b) of the main text employs the Euler--Maruyama scheme~\cite{Kloeden1992} for the dynamics of the mean fields
\begin{align}
\dot{\alpha}&=\biggl\{-i\biggl[\Delta-\sum\limits_{j=1}^Ng_j\text{R}(\beta_j)\biggr]-\frac{\kappa}{2}\biggr\}\alpha+\mathcal{E}_0\mathcal{T}e^{-i\phi_0}+\xi_{\alpha}(t), \nonumber \\
\dot{\beta}_j&=-\bigg(i\Omega_{j}+\frac{\Gamma_j}{2}\bigg)\beta_j+ig_j|\alpha|^2+\xi_{\beta_j}(t),
\label{eq:MeanFieldEvolutionstoch}
\end{align}
where we choose the parameters of the two mechanical modes ($N=2$) as described in the main text, namely $\Omega_1/2\pi=10$ MHz, $\Gamma_1/\Omega_1=0.1$, $g_1=952.717 $ kHz, $\Omega_2/2\pi=11$, $\Gamma_2/\Omega_2=10^{-2}$, $g_2=52.717 $ kHz as well as the optical cavity $\Delta=164.619$ MHz, and $\kappa=88.0211$  MHz. This places the numerical example in the unresolved sideband regime. The Gaussian noise terms we employ are described by their statistical momenta, i.e. their mean $\langle\xi_{s}(t)\rangle=0$ taken to be zero thoughout the analysis and time correlation $\langle\xi_{r}(t)\xi_{s}(t')\rangle=\delta_{rs}\lambda_s\delta(t-t')$ for all $2(N+1)$ variables $r$ and $s$ denoting the real $\text{Re}(z)=\text{R}(z)/2$ and imaginary $\text{Im}(z)=\text{I}(z)/2$ parts of $\alpha$ and $\beta_j$ with the variance of the Gaussian noise $\lambda_s$ gauging the strength of the random forces. Throughout our simulations we employ $\lambda_{\text{Re}(\alpha)}=\lambda_{\text{Im}(\alpha)}=1$ mimicking cavity shot noise as well as noise consistent with the zero point fluctuations of $\beta_1$, described by $\lambda_{\text{Re}(\beta_1)}=\lambda_{\text{Im}(\beta_1)}=1$. The noise in the control oscillator is parametrized by $\lambda_{\text{Re}(\beta_1)}=\lambda_{\text{Im}(\beta_1)}=8001$. We generate an initial condition of the system at the end of the bistable region by evolving the system without noise starting from rest $\alpha(t=-2t_0)=\beta_j(t=-2t_0)=0$ for $t_0=50$ $\mu s$ and constant drive ($\mathcal{E}_0=9107022.675$, $\mathcal{T}_0=(1-i)/2$, $\phi_0=0$). To generate realistic initial conditions, we then repeat the procedure with noise for another $t_0=50$ $\mu s$. After the initial procedure to approach the bistability edge of the system, we then drive with $\mathcal{T}_0=(1-i\mathcal{J}_0(d))/2$, $\mathcal{T}_{\pm 1}=-\mathcal{J}_1(d)$ and switch on the intensity modulation with $d=1.875 \times 10^{-5}$ for $t=200$ $\mu s$. After the modulation has been probed we evolve the system without modulation for another $50$ $\mu s$ to make sure that simulations that were changing steady state have sufficient time to converge and surpass our switching criterion. The bistable state we start from is characterized by a mean number of quanta of $\beta_1$ around 46500 whereas the other state is sustains approximately 79000 oscillator quanta. Thus, switching occurs if the mechanical oscillator quanta of $\beta_1$ surpass 60000 at the end of the simulation. The step size $\delta t=0.0001 \mu s$ throughout every simulation in order to numerically converge. We conducted 50 such runs for modulation with $\Omega_{\text{mod}}=1$ MHz which showed no switching event and another 50 runs with $\Omega_{\text{mod}}=11$ MHz which showed two switching events. This result coincides with the analytic result that intensity modulation at the frequency of the control oscillator at $\Omega_{2}=11$ MHz is resonant and can lead to switching whereas off-resonant optical modulation does not affect the bistable state of $\hat{b}_1$.
We conducted another set of deterministic simulations of
\begin{align}
\dot{\alpha}&=\biggl\{-i\biggl[\Delta-\sum\limits_{j=1}^Ng_j\text{R}(\beta_j)\biggr]-\frac{\kappa}{2}\biggr\}\alpha+\mathcal{E}_0\mathcal{T}, \nonumber \\
\dot{\beta}_j&=-\bigg(i\Omega_{j}+\frac{\Gamma_j}{2}\bigg)\beta_j+ig_j|\alpha|^2+iD\cos(\omega t+\phi_2),
\label{eq:MeanFieldEvolutionstoch}
\end{align}
with $\mathcal{E}_0=9107026.875$, $D=1550$, $d=10^{-4}$ and the system parameters used in the prior simulation. The numerical procedure consists of the initialization process from rest to the parameters at the bistability edge for $t_0= 50$ $\mu s$ with $D=0$ followed by a simulation for $500$ $\mu s$ for the respective phase $\phi_2$ and $\Omega_{\text{mod}}$. The threshold criterion is equivalent to discriminating the steady states by the mean photon number $|\alpha|^2$. Fig 3 (b) of the main text shows that one steady state is characterized by a mean photon number of $3 \times 10^9$ and the other steady state attains a mean photon number of $5 \times 10^9$. Thus our discrimination criterion is to attribute a photon number smaller than $4 \times 10^9$ after the evolution protocol to the initial steady state and a photon number larger than $4 \times 10^9$ to a switching event leading to the phase diagram of Fig. 3 (c) in the main text. The time requirements of the numerical algorithm limit the maximal simulation time per data point leading to fluctuations in the phase diagram because the respective simulations are undergoing the transition but are still below the threshold. 

\section{Modulation depth influence}
\label{sec:AppE}

\begin{figure}[t]
\begin{center}
\includegraphics[width=0.485\textwidth,angle=0]{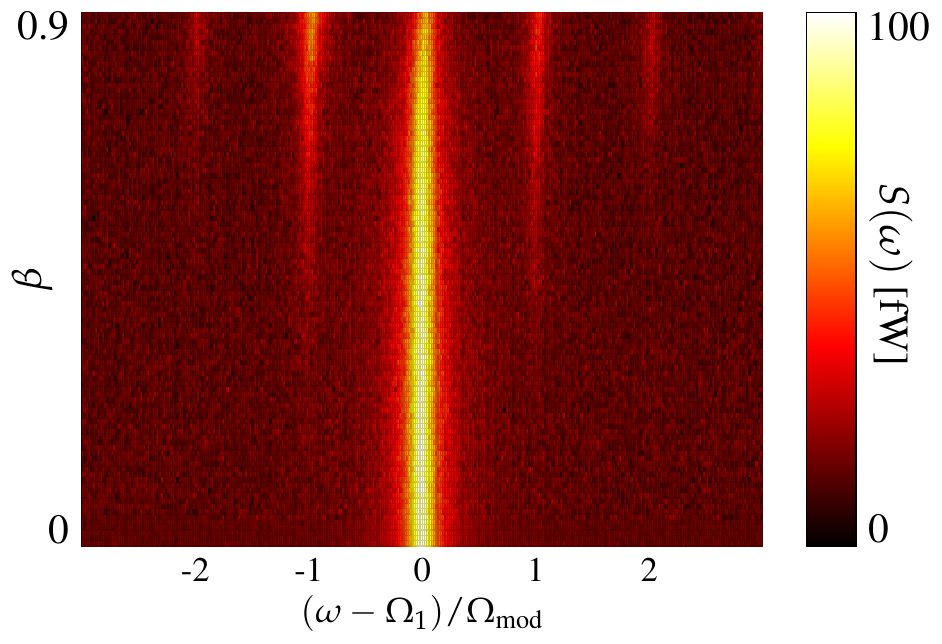}
\caption{Experimental mapping of the fundamental mechanical resonance as a function  of the modulation depth (with $\Omega_\text{mod}=50$ kHz).} 
\label{fig:S2}
\end{center}
\end{figure}
We record the noise spectrum while varying the modulation voltage from 0 to 2 V. The heatmap shown in \fref{S2} evidence the progressive apparition of two pairs of sidebands around the mechanical resonance ($\Omega_1=4.340$). The sidebands start to display imbalance amplitudes around $d=0.75$. The thermo-optically induced imbalance of the modulation sidebands for large modulation depths can be employed for an amplification of the Floquet mechanism. Eq. (8) of the main text implies that an increase of the sideband amplitudes leads to an increased coupling of the Floquet mechanism. We therefore explore the dependence of the amplitude numerically. We employ the same parameters as in Fig. 2 (c) of the main text except for an even larger modulation depth $d=2.0$. These parameters lead to an inverted sideband imbalance as displayed in \fref{S3}. In contrast to the large modulation frequency case, the positive sideband is increased for low modulation frequencies. We therefore find the surprising result that thermo-optical effects can be used to inhibit and to enhance the coupling strength that enables the Floquet control.

\begin{figure}[t]
\begin{center}
\includegraphics[width=0.485\textwidth,angle=0]{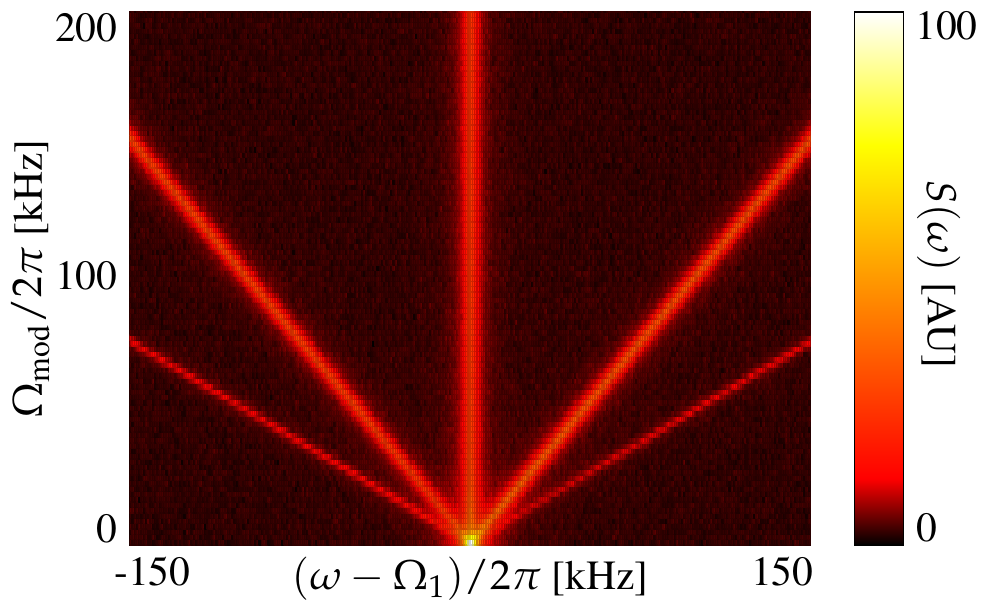}
\caption{Numerical evidence of the inversion of the sideband imbalance for low modulation frequencies with modulation depth $d=2.0$. The increased amplitude of the positive sideband proves that the thermo-optical effect can inhibit or enhance the Floquet control mechanism.} 
\label{fig:S3}
\end{center}
\end{figure}


\begin{thebibliography}{3}%
\makeatletter
\providecommand \@ifxundefined [1]{%
 \@ifx{#1\undefined}
}%
\providecommand \@ifnum [1]{%
 \ifnum #1\expandafter \@firstoftwo
 \else \expandafter \@secondoftwo
 \fi
}%
\providecommand \@ifx [1]{%
 \ifx #1\expandafter \@firstoftwo
 \else \expandafter \@secondoftwo
 \fi
}%
\providecommand \natexlab [1]{#1}%
\providecommand \enquote  [1]{``#1''}%
\providecommand \bibnamefont  [1]{#1}%
\providecommand \bibfnamefont [1]{#1}%
\providecommand \citenamefont [1]{#1}%
\providecommand \href@noop [0]{\@secondoftwo}%
\providecommand \href [0]{\begingroup \@sanitize@url \@href}%
\providecommand \@href[1]{\@@startlink{#1}\@@href}%
\providecommand \@@href[1]{\endgroup#1\@@endlink}%
\providecommand \@sanitize@url [0]{\catcode `\\12\catcode `\$12\catcode
  `\&12\catcode `\#12\catcode `\^12\catcode `\_12\catcode `\%12\relax}%
\providecommand \@@startlink[1]{}%
\providecommand \@@endlink[0]{}%
\providecommand \url  [0]{\begingroup\@sanitize@url \@url }%
\providecommand \@url [1]{\endgroup\@href {#1}{\urlprefix }}%
\providecommand \urlprefix  [0]{URL }%
\providecommand \Eprint [0]{\href }%
\providecommand \doibase [0]{http://dx.doi.org/}%
\providecommand \selectlanguage [0]{\@gobble}%
\providecommand \bibinfo  [0]{\@secondoftwo}%
\providecommand \bibfield  [0]{\@secondoftwo}%
\providecommand \translation [1]{[#1]}%
\providecommand \BibitemOpen [0]{}%
\providecommand \bibitemStop [0]{}%
\providecommand \bibitemNoStop [0]{.\EOS\space}%
\providecommand \EOS [0]{\spacefactor3000\relax}%
\providecommand \BibitemShut  [1]{\csname bibitem#1\endcsname}%
\let\auto@bib@innerbib\@empty

\bibitem [{\citenamefont {Aspelmeyer}\ \emph {et~al.}(2014)\citenamefont
  {Kippenberg},\ and\~\citenamefont
  {Marquardt}}]{Aspelmeyer2014}%
  \BibitemOpen
  \bibfield  {author} {\bibinfo {author} {\bibfnamefont {M.}~\bibnamefont
  {Aspelmeyer}}, \bibinfo {author} {\bibfnamefont {T.~J.}\ \bibnamefont
  {Kippenberg}},\ and\ \bibinfo {author} {\bibfnamefont {F.}~\bibnamefont
  {Marquardt}},\ }\href {\doibase 10.1103/RevModPhys.86.1391} {\bibfield
  {journal} {\bibinfo  {journal} {Rev. Mod. Phys}\ }\textbf {\bibinfo {volume}
  {86}},\ \bibinfo {pages} {1391} (\bibinfo {year} {2014})}\BibitemShut
  {NoStop}%
\bibitem [{\citenamefont {Kippenberg}\ \emph {et~al.}(2005)\citenamefont
  {Kippenberg},~\citenamefont {Rokhsari},~\citenamefont {Carmon},~\citenamefont {Scherer},\ and\~\citenamefont
  {Vahala}}]{Kippenberg2005}%
  \BibitemOpen
  \bibfield  {author} {\bibinfo {author} {\bibfnamefont {T.~J.}~\bibnamefont
  {Kippenberg}}, \bibinfo {author} {\bibfnamefont {H.}\ \bibnamefont
  {Rokhsari}}, \bibinfo {author} {\bibfnamefont {T.}\ \bibnamefont
  {Carmon}}, \bibinfo {author} {\bibfnamefont {A.}\ \bibnamefont
  {Scherer}},\ and\ \bibinfo {author} {\bibfnamefont {K.~J.}~\bibnamefont
  {Vahala}},\ }\href {\doibase 10.1103/PhysRevLett.95.033901} {\bibfield
  {journal} {\bibinfo  {journal} {Phys. Rev. Lett.}\ }\textbf {\bibinfo {volume}
  {95}},\ \bibinfo {pages} {033901} (\bibinfo {year} {2005})}\BibitemShut
  {NoStop}%
\bibitem [{\citenamefont {Marquardt}\ \emph {et~al.}(2006)\citenamefont
  {Marquardt},~\citenamefont {Harris},\ and\~\citenamefont
  {Girvin}}]{Marquardt2006}%
  \BibitemOpen
  \bibfield  {author} {\bibinfo {author} {\bibfnamefont {F.}~\bibnamefont
  {Marquardt}}, \bibinfo {author} {\bibfnamefont {J.~G.~E.}\ \bibnamefont
  {Harris}}, \ and\ \bibinfo {author} {\bibfnamefont {S.~M.}~\bibnamefont
  {Girvin}},\ }\href {\doibase 10.1103/PhysRevLett.96.103901} {\bibfield
  {journal} {\bibinfo  {journal} {Phys. Rev. Lett.}\ }\textbf {\bibinfo {volume}
  {96}},\ \bibinfo {pages} {103901} (\bibinfo {year} {2006})}\BibitemShut
  {NoStop}%
\bibitem [{\citenamefont {Teufel}\ \emph {et~al.}(2011)\citenamefont
  {Donner},~\citenamefont {Li},~\citenamefont {Harlow},~\citenamefont {Allman},~\citenamefont {Cicak},~\citenamefont {Sirois},~\citenamefont {Whittaker},~\citenamefont {Lehnert},\ and\~\citenamefont
  {Simmonds}}]{Teufel2011}%
  \BibitemOpen
  \bibfield  {author} {\bibinfo {author} {\bibfnamefont {J.~D.}~\bibnamefont
  {Teufel}}, \bibinfo {author} {\bibfnamefont {T.}\ \bibnamefont
  {Donner}}, \bibinfo {author} {\bibfnamefont {D.}\ \bibnamefont
  {Li}}, \bibinfo {author} {\bibfnamefont {J.~W.}~ \bibnamefont
  {Harlow}}, \bibinfo {author} {\bibfnamefont {M.~S.}~ \bibnamefont
  {Allman}}, \bibinfo {author} {\bibfnamefont {K.}\ \bibnamefont
  {Cicak}}, \bibinfo {author} {\bibfnamefont {A.~J.}~ \bibnamefont
  {Sirois}}, \bibinfo {author} {\bibfnamefont {J.~D.}~ \bibnamefont
  {Whittaker}}, \bibinfo {author} {\bibfnamefont {K.~W.}~ \bibnamefont
  {Lehnert}},\ and\ \bibinfo {author} {\bibfnamefont {R.~W.}~\bibnamefont
  {Simmonds}},\ }\href {\doibase 10.1038/nature10261} {\bibfield
  {journal} {\bibinfo  {journal} {Nature (London)}\ }\textbf {\bibinfo {volume}
  {475}},\ \bibinfo {pages} {359} (\bibinfo {year} {2011})}\BibitemShut
  {NoStop}%
\bibitem [{\citenamefont {Chan}\ \emph {et~al.}(2011)\citenamefont
  {Alegre},~\citenamefont {Safavi-Naeini},~\citenamefont {Hill},~\citenamefont {Krause},~\citenamefont {Gr\"{o}blacher},~\citenamefont {Aspelmeyer},\ and\~\citenamefont
  {Painter}}]{Chan2011}%
  \BibitemOpen
  \bibfield  {author} {\bibinfo {author} {\bibfnamefont {J.}\ \bibnamefont
  {Chan}}, \bibinfo {author} {\bibfnamefont {T.~P.~M.}\ \bibnamefont
  {Alegre}}, \bibinfo {author} {\bibfnamefont {A.~H.}\ \bibnamefont
  {Safavi-Naeini}}, \bibinfo {author} {\bibfnamefont {J.~T.}\ \bibnamefont
  {Hill}}, \bibinfo {author} {\bibfnamefont {A.}\ \bibnamefont
  {Krause}}, \bibinfo {author} {\bibfnamefont {S.}\ \bibnamefont
  {Gr\"{o}blacher}}, \bibinfo {author} {\bibfnamefont {M.}\ \bibnamefont
  {Aspelmeyer}},\ and\ \bibinfo {author} {\bibfnamefont {O.}\ \bibnamefont
  {Painter}},\ }\href {\doibase 10.1038/nature10261} {\bibfield
  {journal} {\bibinfo  {journal} {Nature (London)}\ }\textbf {\bibinfo {volume}
  {475}},\ \bibinfo {pages} {359} (\bibinfo {year} {2011})}\BibitemShut
  {NoStop}%
\bibitem [{\citenamefont {Ockeloen-Korppi}\ \emph {et~al.}(2018)\citenamefont
  {Damsk\"{a}gg},~\citenamefont {Pirkkalainen},~\citenamefont {Asjad},~\citenamefont {Clerk},~\citenamefont {Massel},~\citenamefont {Woolley},\ and\~\citenamefont
  {Sillanp\"{a}\"{a}}}]{Ockeloen2018}%
  \BibitemOpen
  \bibfield  {author} {\bibinfo {author} {\bibfnamefont {C.~F.}\ \bibnamefont
  {Ockeloen-Korppi}}, \bibinfo {author} {\bibfnamefont {E.}\ \bibnamefont
  {Damsk\"{a}gg}}, \bibinfo {author} {\bibfnamefont {J.~M.}\ \bibnamefont
  {Pirkkalainen}}, \bibinfo {author} {\bibfnamefont {M.}\ \bibnamefont
  {Asjad}}, \bibinfo {author} {\bibfnamefont {A.~A.}\ \bibnamefont
  {Clerk}}, \bibinfo {author} {\bibfnamefont {F.}\ \bibnamefont
  {Massel}}, \bibinfo {author} {\bibfnamefont {M.~J.}\ \bibnamefont
  {Wooley}}, \ and\ \bibinfo {author} {\bibfnamefont {M.~A.}\ \bibnamefont
  {Sillanp\"{a}\"{a}}},\ }\href {\doibase 10.1038/s41586-018-0038-x} {\bibfield
  {journal} {\bibinfo  {journal} {Nature (London)}\ }\textbf {\bibinfo {volume}
  {556}},\ \bibinfo {pages} {478} (\bibinfo {year} {2018})}\BibitemShut
  {NoStop}%
\bibitem [{\citenamefont {Riedinger}\ \emph {et~al.}(2018)\citenamefont
  {Wallucks},~\citenamefont {Marinkovi\'c},~\citenamefont {L\"{o}schnauer},~\citenamefont {Aspelmeyer},~\citenamefont {Hong},\ and\~\citenamefont
  {Gr\"{o}blacher}}]{Riedinger2018}%
  \BibitemOpen
  \bibfield  {author} {\bibinfo {author} {\bibfnamefont {R.}\ \bibnamefont
  {Riedinger}}, \bibinfo {author} {\bibfnamefont {A.}\ \bibnamefont
  {Wallucks}}, \bibinfo {author} {\bibfnamefont {I.}\ \bibnamefont
  {Marinkovi\'c}}, \bibinfo {author} {\bibfnamefont {C.}\ \bibnamefont
  {L\"{o}schnauer}}, \bibinfo {author} {\bibfnamefont {M.}\ \bibnamefont
  {Aspelmeyer}}, \bibinfo {author} {\bibfnamefont {S.}\ \bibnamefont
  {Hong}},\ and\ \bibinfo {author} {\bibfnamefont {S.}\ \bibnamefont
  {Gr\"{o}blacher}},\ }\href {\doibase 10.1038/s41586-018-0036-z} {\bibfield
  {journal} {\bibinfo  {journal} {Nature (London)}\ }\textbf {\bibinfo {volume}
  {556}},\ \bibinfo {pages} {473} (\bibinfo {year} {2018})}\BibitemShut
  {NoStop}%
\bibitem [{\citenamefont {Heinrich}\ \emph {et~al.}(2011)\citenamefont
  {Heinrich},~\citenamefont {Ludwig},\citenamefont {Qian},\citenamefont {Kubala},\ and\~\citenamefont
  {Marquardt}}]{Heinrich2011}%
  \BibitemOpen
  \bibfield  {author} {\bibinfo {author} {\bibfnamefont {G.}~\bibnamefont
  {Heinrich}}, \bibinfo {author} {\bibfnamefont {M.}~\bibnamefont
  {Ludwig}}, \bibinfo {author} {\bibfnamefont {J.}~\bibnamefont
  {Qian}}, \bibinfo {author} {\bibfnamefont {B.}\ \bibnamefont
  {Kubala}},\ and\ \bibinfo {author} {\bibfnamefont {F.}~\bibnamefont
  {Marquardt}},\ }\href {\doibase 10.1103/PhysRevLett.107.043603} {\bibfield
  {journal} {\bibinfo  {journal} {Phys. Rev. Lett.}\ }\textbf {\bibinfo {volume}
  {107}},\ \bibinfo {pages} {043603} (\bibinfo {year} {2011})}\BibitemShut
  {NoStop}%
\bibitem [{\citenamefont {Lauter}\ \emph {et~al.}(2015)\citenamefont
  {Lauter},~\citenamefont {Brendel},\citenamefont {Habraken},\ and\~\citenamefont
  {Marquardt}}]{Lauter2015}%
  \BibitemOpen
  \bibfield  {author} {\bibinfo {author} {\bibfnamefont {R.}~\bibnamefont
  {Lauter}}, \bibinfo {author} {\bibfnamefont {C.}~\bibnamefont
  {Brendel}}, \bibinfo {author} {\bibfnamefont {S.~J.~M.}~\bibnamefont
  {Habraken}},\ and\ \bibinfo {author} {\bibfnamefont {F.}~\bibnamefont
  {Marquardt}},\ }\href {\doibase 10.1103/PhysRevE.92.012902} {\bibfield
  {journal} {\bibinfo  {journal} {Phys. Rev. E}\ }\textbf {\bibinfo {volume}
  {92}},\ \bibinfo {pages} {012902} (\bibinfo {year} {2015})}\BibitemShut
  {NoStop}%
\bibitem [{\citenamefont {Lauter}\ \emph {et~al.}(2015)\citenamefont
  {Lauter},~\citenamefont {Mitra},\ and\~\citenamefont
  {Marquardt}}]{Lauter2017}%
  \BibitemOpen
  \bibfield  {author} {\bibinfo {author} {\bibfnamefont {R.}~\bibnamefont
  {Lauter}}, \bibinfo {author} {\bibfnamefont {A.}~\bibnamefont
  {Mitra}},\ and\ \bibinfo {author} {\bibfnamefont {F.}~\bibnamefont
  {Marquardt}},\ }\href {\doibase 10.1103/PhysRevE.96.012220} {\bibfield
  {journal} {\bibinfo  {journal} {Phys. Rev. E}\ }\textbf {\bibinfo {volume}
  {96}},\ \bibinfo {pages} {012220} (\bibinfo {year} {2017})}\BibitemShut
  {NoStop}%
\bibitem [{\citenamefont {Holmes}\ \emph {et~al.}(2012)\citenamefont
  {Meaney},\ and\~\citenamefont
  {Milburn}}]{Holmes2012}%
  \BibitemOpen
  \bibfield  {author} {\bibinfo {author} {\bibfnamefont {C.~A.}\ \bibnamefont
  {Holmes}}, \bibinfo {author} {\bibfnamefont {C.~P.}\ \bibnamefont
  {Meaney}},\ and\ \bibinfo {author} {\bibfnamefont {G.~J.}\ \bibnamefont
  {Milburn}},\ }\href {\doibase 10.1103/PhysRevE.85.066203} {\bibfield
  {journal} {\bibinfo  {journal} {Phys. Rev. E}\ }\textbf {\bibinfo {volume}
  {85}},\ \bibinfo {pages} {066203} (\bibinfo {year} {2012})}\BibitemShut
  {NoStop}%
\bibitem [{\citenamefont {L\"{o}rch}\ \emph {et~al.}(2017)\citenamefont
  {Nigg},~\citenamefont {Nunnenkamp},~\citenamefont {Tiwari},\ and\~\citenamefont
  {Bruder}}]{Loerch2017}%
  \BibitemOpen
  \bibfield  {author} {\bibinfo {author} {\bibfnamefont {N.}~\bibnamefont
  {L\"{o}rch}}, \bibinfo {author} {\bibfnamefont {S.~E.}~\bibnamefont
  {Nigg}}, \bibinfo {author} {\bibfnamefont {A.}~\bibnamefont
  {Nunnenkamp}}, \bibinfo {author} {\bibfnamefont {R.~P.}~\bibnamefont
  {Tiwari}}, \ and\ \bibinfo {author} {\bibfnamefont {C.}~\bibnamefont
  {Bruder}},\ }\href {\doibase 10.1103/PhysRevLett.118.243602} {\bibfield
  {journal} {\bibinfo  {journal} {Phys. Rev. Lett}\ }\textbf {\bibinfo {volume}
  {118}},\ \bibinfo {pages} {243602} (\bibinfo {year} {2017})}\BibitemShut
  {NoStop}%
\bibitem [{\citenamefont {Amitai}\ \emph {et~al.}(2017)\citenamefont
  {L\"{o}rch},~\citenamefont {Nunnenkamp},~\citenamefont {Walter},\ and\~\citenamefont
  {Bruder}}]{Amitai2017}%
  \BibitemOpen
  \bibfield  {author} {\bibinfo {author} {\bibfnamefont {E.}~\bibnamefont
  {Amitai}}, \bibinfo {author} {\bibfnamefont {N.}~\bibnamefont
  {L\"{o}rch}}, \bibinfo {author} {\bibfnamefont {A.}~\bibnamefont
  {Nunnenkamp}}, \bibinfo {author} {\bibfnamefont {S.}~\bibnamefont
  {Walter}}, \ and\ \bibinfo {author} {\bibfnamefont {C.}~\bibnamefont
  {Bruder}},\ }\href {\doibase 10.1103/PhysRevA.95.053858} {\bibfield
  {journal} {\bibinfo  {journal} {Phys. Rev. A}\ }\textbf {\bibinfo {volume}
  {95}},\ \bibinfo {pages} {053858} (\bibinfo {year} {2017})}\BibitemShut
  {NoStop}%
\bibitem [{\citenamefont {Zhang}\ \emph {et~al.}(2012)\citenamefont
  {Wiederhecker},~\citenamefont {Manipatruni},~\citenamefont {Barnard},~\citenamefont {McEuen},\ and\~\citenamefont
  {Lipson}}]{Lipson2012}%
  \BibitemOpen
  \bibfield  {author} {\bibinfo {author} {\bibfnamefont {M.}~\bibnamefont
  {Zhang}}, \bibinfo {author} {\bibfnamefont {G.S.}~\bibnamefont
  {Wiederhecker}}, \bibinfo {author} {\bibfnamefont {S.}~\bibnamefont
  {Manipatruni}}, \bibinfo {author} {\bibfnamefont {A.}~\bibnamefont
  {Barnard}}, \bibinfo {author} {\bibfnamefont {P.}\ \bibnamefont
  {McEuen}}, \ and\ \bibinfo {author} {\bibfnamefont {M.}~\bibnamefont
  {Lipson}},\ }\href {\doibase 10.1103/PhysRevLett.109.233906} {\bibfield
  {journal} {\bibinfo  {journal} {Phys. Rev. Lett.}\ }\textbf {\bibinfo {volume}
  {109}},\ \bibinfo {pages} {233906} (\bibinfo {year} {2012})}\BibitemShut
  {NoStop}%
\bibitem [{\citenamefont {Zhang}\ \emph {et~al.}(2015)\citenamefont
  {Zhang},~\citenamefont {Wiederhecker},\citenamefont {Cardenas},\ and\~\citenamefont
  {Lipson}}]{Lipson2015}%
    \BibitemOpen   
  \bibfield  {author} {\bibinfo {author} {\bibfnamefont {M.}~\bibnamefont
  {Zhang}}, \bibinfo {author} {\bibfnamefont {S.}~\bibnamefont
  {Shah}}, \bibinfo {author} {\bibfnamefont {J.}~\bibnamefont
  {Cardenas}},\ and\ \bibinfo {author} {\bibfnamefont {M.}~\bibnamefont
  {Lipson}},\ }\href {\doibase 10.1103/PhysRevLett.115.163902} {\bibfield
  {journal} {\bibinfo  {journal} {Phys. Rev. Lett.}\ }\textbf {\bibinfo {volume}
  {115}},\ \bibinfo {pages} {163902} (\bibinfo {year} {2015})}\BibitemShut
  {NoStop}%
\bibitem [{\citenamefont {Colombano}\ \emph {et~al.}(2019)\citenamefont
  {Arregui},~\citenamefont {Capuj},~\citenamefont {Pitanti}, 
 ~\citenamefont {Maire},~\citenamefont {Griol},~\citenamefont 
  {Garrido},~\citenamefont {Martinez},\citenamefont {Sotomayor-Torres},\ 
  and\~\citenamefont{Navarro-Urrios}}]{Colombano2019}%
    \BibitemOpen   
  \bibfield  {author} {\bibinfo {author} {\bibfnamefont {M.~F.}~\bibnamefont
  {Colombano}}, \bibinfo {author} {\bibfnamefont {G.}~\bibnamefont
  {Arregui}}, \bibinfo {author} {\bibfnamefont {N.~E.}~\bibnamefont
  {Capuj}}, \bibinfo {author} {\bibfnamefont {A.}~\bibnamefont
  {Pitanti}}, \bibinfo {author} {\bibfnamefont {J.}~\bibnamefont
  {Maire}}, \bibinfo {author} {\bibfnamefont {A.}~\bibnamefont
  {Griol}}, \bibinfo {author} {\bibfnamefont {B.}~\bibnamefont
  {Garrido}}, \bibinfo {author} {\bibfnamefont {A.}~\bibnamefont
  {Martinez}}, \bibinfo {author} {\bibfnamefont {C.~M.}~\bibnamefont
  {Sotomayor-Torres}},\ and\ \bibinfo {author} {\bibfnamefont {D.}~\bibnamefont
  {Navarros-Urrios}},\ }\href {\doibase 10.1103/PhysRevLett.123.017402} {\bibfield
  {journal} {\bibinfo  {journal} {Phys. Rev. Lett.}\ }\textbf {\bibinfo {volume}
  {123}},\ \bibinfo {pages} {017402} (\bibinfo {year} {2019})}\BibitemShut
  {NoStop}%
\bibitem [{\citenamefont {Pelka}\ \emph {et~al.}(2020)\citenamefont
  {Peano},\ 
  and\~\citenamefont{Xuereb}}]{Pelka2020}%
    \BibitemOpen   
  \bibfield  {author} {\bibinfo {author} {\bibfnamefont {K.}~\bibnamefont
  {Pelka}}, \bibinfo {author} {\bibfnamefont {V.}~\bibnamefont
  {Peano}},\ and\ \bibinfo {author} {\bibfnamefont {A.}~\bibnamefont
  {Xuereb}},\ }\href {\doibase 10.1103/PhysRevResearch.2.013201} {\bibfield
  {journal} {\bibinfo  {journal} {Phys. Rev. Research}\ }\textbf {\bibinfo {volume}
  {123}},\ \bibinfo {pages} {017402} (\bibinfo {year} {2019})}\BibitemShut
  {NoStop}%
\bibitem [{\citenamefont {Madiot}\ \emph {et~al.}(2020)\citenamefont
  {Correia},\citenamefont {Barbay},\ 
  and\~\citenamefont{Braive}}]{Madiot2020}%
    \BibitemOpen   
  \bibfield  {author} {\bibinfo {author} {\bibfnamefont {G.}~\bibnamefont
  {Madiot}}, \bibinfo {author} {\bibfnamefont {F.}~\bibnamefont
  {Correia}}, \bibinfo {author} {\bibfnamefont {S.}~\bibnamefont
  {Barbay}},\ and\ \bibinfo {author} {\bibfnamefont {R.}~\bibnamefont
  {Braive}},\ }\href {https://arxiv.org/abs/2005.08896} {\bibfield
  {journal} {\bibinfo  {journal} {arXiv:2005.08896}}\textbf {\bibinfo {volume}
  {}}, \bibinfo {pages} {} (\bibinfo {year} {2020})}\BibitemShut
  {NoStop}%
\bibitem [{\citenamefont {Bernier}\ \emph {et~al.}(2017)\citenamefont
  {T\'{o}th},~\citenamefont {Koottandavida},~\citenamefont {Ioannou},~\citenamefont {Malz},~\citenamefont {Nunnenkamp},~\citenamefont {Feofanov},\ and\~\citenamefont
  {Kippenberg}}]{Bernier2017}%
  \BibitemOpen
  \bibfield  {author} {\bibinfo {author} {\bibfnamefont {N.~R.}\ \bibnamefont
  {Bernier}}, \bibinfo {author} {\bibfnamefont {L.~D.}\ \bibnamefont
  {T\'{o}th}}, \bibinfo {author} {\bibfnamefont {A.}\ \bibnamefont
  {Koottandavida}}, \bibinfo {author} {\bibfnamefont {M.~D.}\ \bibnamefont
  {Ioannu}}, \bibinfo {author} {\bibfnamefont {D.}\ \bibnamefont
  {Malz}}, \bibinfo {author} {\bibfnamefont {A.}\ \bibnamefont
  {Nunnenkamp}}, \bibinfo {author} {\bibfnamefont {A.~K.}\ \bibnamefont
  {Feofanov}},\ and\ \bibinfo {author} {\bibfnamefont {T.~J.}\ \bibnamefont
  {Kippenberg}},\ }\href {\doibase 10.1038/s41467-017-00447-1} {\bibfield
  {journal} {\bibinfo  {journal} {Nat. Commun.}\ }\textbf {\bibinfo {volume}
  {8}},\ \bibinfo {pages} {604} (\bibinfo {year} {2017})}\BibitemShut
  {NoStop}%
\bibitem [{\citenamefont {Malz}\ \emph {et~al.}(2018)\citenamefont
  {T\'{o}th},~\citenamefont {Bernier},~\citenamefont {Feofanov},~\citenamefont {Kippenberg},\ and\~\citenamefont
  {Nunnenkamp}}]{Malz2018}%
  \BibitemOpen
  \bibfield  {author} { \bibinfo {author} {\bibfnamefont {D.}\ \bibnamefont
  {Malz}}, \bibinfo {author} {\bibfnamefont {L.~D.}\ \bibnamefont
  {T\'{o}th}}, \bibinfo {author} {\bibfnamefont {N.~R.}\ \bibnamefont
  {Bernier}},  \bibinfo {author} {\bibfnamefont {A.~K.}\ \bibnamefont
  {Feofanov}}, \bibinfo {author} {\bibfnamefont {T.~J.}\ \bibnamefont
  {Kippenberg}},\ and\ \bibinfo {author} {\bibfnamefont {A.}\ \bibnamefont
  {Nunnenkamp}},\ }\href {\doibase 10.1103/PhysRevLett.120.023601} {\bibfield
  {journal} {\bibinfo  {journal} {Phys. Rev. Lett.}\ }\textbf {\bibinfo {volume}
  {120}},\ \bibinfo {pages} {023601} (\bibinfo {year} {2018})}\BibitemShut
  {NoStop}%
\bibitem [{\citenamefont {Barzanjeh}\ \emph {et~al.}(2017)\citenamefont
  {Wulf},~\citenamefont {Peruzzo},~\citenamefont {Kalaee},~\citenamefont {Dieterle},~\citenamefont {Painter},\ and\~\citenamefont
  {Fink}}]{Barzanjeh2017}%
  \BibitemOpen
  \bibfield  {author} {\bibinfo {author} {\bibfnamefont {S.}\ \bibnamefont
  {Barzanjeh}}, \bibinfo {author} {\bibfnamefont {M.}\ \bibnamefont
  {Wulf}}, \bibinfo {author} {\bibfnamefont {M.}\ \bibnamefont
  {Peruzzo}}, \bibinfo {author} {\bibfnamefont {M.}\ \bibnamefont
  {Kalaee}}, \bibinfo {author} {\bibfnamefont {P.~B.}\ \bibnamefont
  {Dieterle}}, \bibinfo {author} {\bibfnamefont {O.}\ \bibnamefont
  {Painter}},\ and\ \bibinfo {author} {\bibfnamefont {J.~M.}\ \bibnamefont
  {Fink}},\ }\href {\doibase 10.1038/s41467-017-01304-x} {\bibfield
  {journal} {\bibinfo  {journal} {Nat. Commun.}\ }\textbf {\bibinfo {volume}
  {8}},\ \bibinfo {pages} {953} (\bibinfo {year} {2017})}\BibitemShut
  {NoStop}%
\bibitem [{\citenamefont {Mercier de L\'epinay}\ \emph {et~al.}(2019)\citenamefont
  {Genes},\ and\~\citenamefont
  {Dantan}}]{Mercier2019}%
  \BibitemOpen
  \bibfield  {author} {\bibinfo {author} {\bibfnamefont {L.}~\bibnamefont
  {Mercier de L\'epinay}}, \bibinfo {author} {\bibfnamefont {E.}\ \bibnamefont
  {Damsk\"agg}}, \bibinfo {author} {\bibfnamefont {C.~F.}\ \bibnamefont
  {Ockeloen-Korppi}},\ and\ \bibinfo {author} {\bibfnamefont {M.~A.}~\bibnamefont
  {Sillanp\"a\"a}},\ }\href {\doibase 10.1103/PhysRevApplied.11.034027} {\bibfield
  {journal} {\bibinfo  {journal} {Phys. Rev. Applied}\ }\textbf {\bibinfo {volume}
  {11}},\ \bibinfo {pages} {034027} (\bibinfo {year} {2019})}\BibitemShut
  {NoStop}%
\bibitem [{\citenamefont {Andrews}\ \emph {et~al.}(2014)\citenamefont
  {Peterson},~\citenamefont {Purdy},~\citenamefont {Cicak},~\citenamefont {Simmonds},~\citenamefont {Regal},\ and\~\citenamefont
  {Lehnert}}]{Andrews2014}%
  \BibitemOpen
  \bibfield  {author} {\bibinfo {author} {\bibfnamefont {R.~W.}~\bibnamefont
  {Andrews}}, \bibinfo {author} {\bibfnamefont {T.~P.}\ \bibnamefont
  {Purdy}}, \bibinfo {author} {\bibfnamefont {K.}\ \bibnamefont
  {Cicak}}, \bibinfo {author} {\bibfnamefont {R.~W.}\ \bibnamefont
  {Simmonds}}, \bibinfo {author} {\bibfnamefont {C.~A.}\ \bibnamefont
  {Regal}},\ and\ \bibinfo {author} {\bibfnamefont {K.~W.}~\bibnamefont
  {Lehnert}},\ }\href {\doibase 10.1038/NPHYS2911} {\bibfield
  {journal} {\bibinfo  {journal} {Nat. Phys.}\ }\textbf {\bibinfo {volume}
  {11}},\ \bibinfo {pages} {034027} (\bibinfo {year} {2019})}\BibitemShut
  {NoStop}%
\bibitem [{\citenamefont {Dorsel}\ \emph {et~al.}(1983)\citenamefont
  {McCullen}, \citenamefont{Meystre}, \citenamefont{Vignes}, \ and\~\citenamefont
  {Walther}}]{Dorsel1983}%
  \BibitemOpen
  \bibfield  {author} {\bibinfo {author} {\bibfnamefont {A.}\ \bibnamefont
  {Dorsel}}, \bibinfo {author} {\bibfnamefont {J.~D.}\ \bibnamefont
  {McCullen}}, \bibinfo {author} {\bibfnamefont {P.}\ \bibnamefont
  {Meystre}}, \bibinfo {author} {\bibfnamefont {E.}\ \bibnamefont
  {Vignes}},\ and\ \bibinfo {author} {\bibfnamefont {H.}\ \bibnamefont
  {Walther}},\ }\href {\doibase 10.1103/PhysRevLett.51.1550} {\bibfield
  {journal} {\bibinfo  {journal} {Phys. Rev. Lett.}\ }\textbf {\bibinfo {volume}
  {51}},\ \bibinfo {pages} {1550} (\bibinfo {year} {1983})}\BibitemShut
  {NoStop}%
\bibitem [{\citenamefont {Ghobadi}\ \emph {et~al.}(1983)\citenamefont
  {Bahrampour}, \ and\~\citenamefont
  {Simon}}]{Ghobadi2011}%
  \BibitemOpen
  \bibfield  {author} {\bibinfo {author} {\bibfnamefont {R.}\ \bibnamefont
  {Ghobadi}}, \bibinfo {author} {\bibfnamefont {A.~R.}\ \bibnamefont
  {Bahrampour}},\ and\ \bibinfo {author} {\bibfnamefont {C.}\ \bibnamefont
  {Simon}},\ }\href {\doibase 10.1103/PhysRevA.84.033846} {\bibfield
  {journal} {\bibinfo  {journal} {Phys. Rev. A}\ }\textbf {\bibinfo {volume}
  {84}},\ \bibinfo {pages} {033846} (\bibinfo {year} {2011})}\BibitemShut
  {NoStop}%
\bibitem [{\citenamefont {Badzey}\ \emph {et~al.}(2004)\citenamefont{Zolfagharkhani}, \citenamefont{Gaidarzhy}, \ and\~\citenamefont
  {Mohanty}}]{Badzey2004}%
  \BibitemOpen
  \bibfield  {author} {\bibinfo {author} {\bibfnamefont {R.~L.}\ \bibnamefont
  {Badzey}}, \bibinfo {author} {\bibfnamefont {G.}\ \bibnamefont
  {Zolfagharkhani}}, \bibinfo {author} {\bibfnamefont {A.}\ \bibnamefont
  {Gaidarzhy}}\ and\ \bibinfo {author} {\bibfnamefont {P.}\ \bibnamefont
  {Mohanty}},\ }\href {\doibase 10.1063/1.1808507} {\bibfield
  {journal} {\bibinfo  {journal} {Appl. Phys. Lett.}\ }\textbf {\bibinfo {volume}
  {85}},\ \bibinfo {pages} {3587} (\bibinfo {year} {2004})}\BibitemShut
  {NoStop}%
\bibitem [{\citenamefont {Maboob}\ \emph {et~al.}(2008)\ and\~\citenamefont
  {Yamaguchi}}]{Maboob2008}%
  \BibitemOpen
  \bibfield  {author} {\bibinfo {author} {\bibfnamefont {I.}~\bibnamefont
  {Maboob}}, \bibinfo {author} {\bibfnamefont {H.}~\bibnamefont
  {Yamaguchi}},\ }\href {\doibase 10.1038/nnano.2008.84} {\bibfield
  {journal} {\bibinfo  {journal} {Nat. Nanotechnol.}\ }\textbf {\bibinfo {volume}
  {3}},\ \bibinfo {pages} {275} (\bibinfo {year} {2008})}\BibitemShut
  {NoStop}%
\bibitem [{\citenamefont {Bagheri}\ \emph {et~al.}(2011)\citenamefont
  {Poot}, \citenamefont {Poot}, \citenamefont {Li}, \citenamefont
  {Pernice}, \ and\~\citenamefont
  {Tang}}]{Bagheri2011}%
  \BibitemOpen
  \bibfield  {author} {\bibinfo {author} {\bibfnamefont {M.}~\bibnamefont
  {Bagheri}}, \bibinfo {author} {\bibfnamefont {M.}~\bibnamefont
  {Poot}}, \bibinfo {author} {\bibfnamefont {M.}~\bibnamefont
  {Li}}, \bibinfo {author} {\bibfnamefont {W.~P.~H.}~\bibnamefont
  {Pernice}},\ and\ \bibinfo {author} {\bibfnamefont {H.~X.}~\bibnamefont
  {Tang}},\ }\href {\doibase 10.1038/nnano.2011.180} {\bibfield
  {journal} {\bibinfo  {journal} {Nat. Nanotechnol.}\ }\textbf {\bibinfo {volume}
  {6}},\ \bibinfo {pages} {726} (\bibinfo {year} {2011})}\BibitemShut
  {NoStop}%
\bibitem [{\citenamefont {Malz}\ \emph {et~al.}(2016),\ and\~\citenamefont
  {Nunnenkamp}}]{Malz2016}%
  \BibitemOpen
  \bibfield  {author} {\bibinfo {author} {\bibfnamefont {D.}~\bibnamefont
  {Malz}},\ and\ \bibinfo {author} {\bibfnamefont {A.}~\bibnamefont
  {Nunnenkamp}},\ }\href {\doibase 10.1103/PhysRevA.94.023803} {\bibfield
  {journal} {\bibinfo  {journal} {Phys. Rev. A}\ }\textbf {\bibinfo {volume}
  {94}},\ \bibinfo {pages} {023803} (\bibinfo {year} {2016})}\BibitemShut
  {NoStop}%
\bibitem [{\citenamefont {Pietik\"{a}inen}\ \emph {et~al.}(2020)\citenamefont
  {Cernotik}, \ and\~\citenamefont
  {Filip}}]{Pietikainen2020}%
  \BibitemOpen
  \bibfield  {author} {\bibinfo {author} {\bibfnamefont {I.}\ \bibnamefont
  {Pietik\"{a}inen}}, \bibinfo {author} {\bibfnamefont {O.}\ \bibnamefont
  {$\breve{C}$ernotik}},\ and\ \bibinfo {author} {\bibfnamefont {R.}\ \bibnamefont
  {Filip}},\ }\href {\doibase 10.1088/1367-2630/ab8cab} {\bibfield
  {journal} {\bibinfo  {journal} {New J. Phys.}\ }\textbf {\bibinfo {volume}
  {22}},\ \bibinfo {pages} {063019} (\bibinfo {year} {2020})}\BibitemShut
  {NoStop}%
\bibitem [{\citenamefont {Xu}\ \emph {et~al.}(2019)\citenamefont
  {Jiang},\ and\~\citenamefont
  {Harris}}]{Xu2019}%
  \BibitemOpen
  \bibfield  {author} {\bibinfo {author} {\bibfnamefont {H.}~\bibnamefont
  {Xu}}, \bibinfo {author} {\bibfnamefont {A.~A.}~\bibnamefont
  {Clerk}},\ and\ \bibinfo {author} {\bibfnamefont {J.~G.~E.}~\bibnamefont
  {Harris}},\ }\href {\doibase 10.1038/s41586-019-1061-2} {\bibfield
  {journal} {\bibinfo  {journal} {Nature}\ }\textbf {\bibinfo {volume}
  {568}},\ \bibinfo {pages} {65} (\bibinfo {year} {2019})}\BibitemShut
  {NoStop}
\bibitem [{\citenamefont {Peano}\ \emph {et~al.}(2015)\citenamefont
  {Brendel},\~\citenamefont
  {Schmidt},\ and\~\citenamefont
  {Marquardt}}]{Peano2015}%
  \BibitemOpen
  \bibfield  {author} {\bibinfo {author} {\bibfnamefont {V.}~\bibnamefont
  {Peano}}, \bibinfo {author} {\bibfnamefont {C.}\ \bibnamefont
  {Brendel}}, \bibinfo {author} {\bibfnamefont {M.}\ \bibnamefont
  {Schmidt}},\ and\ \bibinfo {author} {\bibfnamefont {F.}~\bibnamefont
  {Marquardt}},\ }\href {\doibase 10.1103/PhysRevX.5.031011} {\bibfield
  {journal} {\bibinfo  {journal} {Phys. Rev. X}\ }\textbf {\bibinfo {volume}
  {5}},\ \bibinfo {pages} {031011} (\bibinfo {year} {2015})}\BibitemShut
  {NoStop}%
\bibitem [{\citenamefont {Walter}\ \emph {et~al.}(2016)\citenamefont \ and\~\citenamefont
  {Marquardt}}]{Walter2016}%
  \BibitemOpen
  \bibfield  {author} {\bibinfo {author} {\bibfnamefont {S.}~\bibnamefont
  {Walter}},\ and\ \bibinfo {author} {\bibfnamefont {F.}~\bibnamefont
  {Marquardt}},\ }\href {\doibase 10.1088/1367-2630/18/11/113029} {\bibfield
  {journal} {\bibinfo  {journal} {New J. Phys.}\ }\textbf {\bibinfo {volume}
  {18}},\ \bibinfo {pages} {113029} (\bibinfo {year} {2016})}\BibitemShut
  {NoStop}%
\bibitem [{\citenamefont {Mathew}\ \emph {et~al.}(2018)\citenamefont
  {del Pino}, \ and\~\citenamefont
  {Verhagen}}]{Mathew2018}%
  \BibitemOpen
  \bibfield  {author} {\bibinfo {author} {\bibfnamefont {J.~P.}~\bibnamefont
  {Mathew}}, \bibinfo {author} {\bibfnamefont {J.}~\bibnamefont
  {del Pino}},\ and\ \bibinfo {author} {\bibfnamefont {E.}~\bibnamefont
  {Verhagen}},\ }\href {\doibase 10.1038/s41565-019-0630-8} {\bibfield
  {journal} {\bibinfo  {journal} {Nat. Nanotechnol.}\ }\textbf {\bibinfo {volume}
  {15}},\ \bibinfo {pages} {198} (\bibinfo {year} {2020})}\BibitemShut
  {NoStop}%
\bibitem [{\citenamefont {Weaver}\ \emph {et~al.}(2017)\citenamefont
  {Buters},\citenamefont {Luna},\citenamefont {Eerkens},\citenamefont {Heeck},\citenamefont {de Man},\ and\~\citenamefont
  {Bouwmeester}}]{Weaver2017}%
  \BibitemOpen
  \bibfield  {author} {\bibinfo {author} {\bibfnamefont {M.~J.}~\bibnamefont
  {Weaver}}, \bibinfo {author} {\bibfnamefont {F.}~\bibnamefont
  {Buters}}, \bibinfo {author} {\bibfnamefont {F.}~\bibnamefont
  {Luna}}, \bibinfo {author} {\bibfnamefont {H.}~\bibnamefont
  {Eerkens}}, \bibinfo {author} {\bibfnamefont {K.}~\bibnamefont
  {Heeck}}, \bibinfo {author} {\bibfnamefont {S.}~\bibnamefont
  {de Man}},\ and\ \bibinfo {author} {\bibfnamefont {J.~G.~E.}~\bibnamefont
  {Harris}},\ }\href {\doibase 10.1038/s41467-017-00968-9} {\bibfield
  {journal} {\bibinfo  {journal} {Nat. Commun.}\ }\textbf {\bibinfo {volume}
  {568}},\ \bibinfo {pages} {65} (\bibinfo {year} {2019})}\BibitemShut
  {NoStop}
\bibitem [{\citenamefont {Mercad\'{e}}\ \emph {et~al.}(2021)\citenamefont
  {Pelka}, \citenamefont{Burgwal}, \citenamefont{Xuereb}, \citenamefont{Mart\'{i}nez}, \ and\~\citenamefont
  {Verhagen}}]{Mercade2021}%
  \BibitemOpen
  \bibfield  {author} {\bibinfo {author} {\bibfnamefont {L.}\ \bibnamefont
  {Mercad\'{e}}}, \bibinfo {author} {\bibfnamefont {K.}\ \bibnamefont
  {Pelka}}, \bibinfo {author} {\bibfnamefont {R.}\ \bibnamefont
  {Burgwal}}, \bibinfo {author} {\bibfnamefont {A.}\ \bibnamefont
  {Xuereb}}, \bibinfo {author} {\bibfnamefont {A.}\ \bibnamefont
  {Mart\'{i}nez}},\ and\ \bibinfo {author} {\bibfnamefont {E.}\ \bibnamefont
  {Verhagen}},\ }\href {https://arxiv.org/abs/2101.10788} {\bibfield
  {journal} {\bibinfo  {journal} {arXiv:2101.10788}}\textbf {\bibinfo {volume}
  {}}, \bibinfo {pages} {} (\bibinfo {year} {2021})}\BibitemShut
  {NoStop}%
\bibitem [{\citenamefont {Eichenfield}\ \emph {et~al.}(2019)\citenamefont
  {Kamacho},~\citenamefont {Chan},~\citenamefont {Vahala},\ and\~\citenamefont
  {Painter}}]{Eichenfield2009}%
  \BibitemOpen
  \bibfield  {author} {\bibinfo {author} {\bibfnamefont {M.}~\bibnamefont
  {Eichenfield}}, \bibinfo {author} {\bibfnamefont {R.}~\bibnamefont
  {Kamacho}}, \bibinfo {author} {\bibfnamefont {J.}~\bibnamefont
  {Chan}}, \bibinfo {author} {\bibfnamefont {K.~J.}~\bibnamefont
  {Vahala}},\ and\ \bibinfo {author} {\bibfnamefont {O.}~\bibnamefont
  {Painter}},\ }\href {\doibase 10.1038/nature08061} {\bibfield
  {journal} {\bibinfo  {journal} {Nature}\ }\textbf {\bibinfo {volume}
  {459}},\ \bibinfo {pages} {550} (\bibinfo {year} {2009})}\BibitemShut
  {NoStop}
\bibitem [{\citenamefont {Verhagen}\ \emph {et~al.}(2012)\citenamefont
  {Del\'eglise},~\citenamefont {Weis},~\citenamefont {Schliesser},\ and\~\citenamefont
  {Kippenberg}}]{Verhagen2012}%
  \BibitemOpen
  \bibfield  {author} {\bibinfo {author} {\bibfnamefont {E.}~\bibnamefont
  {Verhagen}}, \bibinfo {author} {\bibfnamefont {S.}~\bibnamefont
  {Del\'eglise}}, \bibinfo {author} {\bibfnamefont {S.}~\bibnamefont
  {Weis}}, \bibinfo {author} {\bibfnamefont {A.}~\bibnamefont
  {Schliesser}},\ and\ \bibinfo {author} {\bibfnamefont {T.~J.}~\bibnamefont
  {Kippenberg}},\ }\href {\doibase 10.1038/nature10787} {\bibfield
  {journal} {\bibinfo  {journal} {Nature}\ }\textbf {\bibinfo {volume}
  {482}},\ \bibinfo {pages} {63} (\bibinfo {year} {2012})}\BibitemShut
  {NoStop}
\bibitem [{\citenamefont {Li}\ \emph {et~al.}(2014)\citenamefont
  {Diddams},\ and\~\citenamefont {Vahala}}]{Li2014}%
  \BibitemOpen
  \bibfield  {author} {\bibinfo {author} {\bibfnamefont {J.}~\bibnamefont
  {Li}}, \bibinfo {author} {\bibfnamefont {S.}~\bibnamefont
  {Diddams}},\ and\ \bibinfo {author} {\bibfnamefont {K.~J.}~\bibnamefont
  {Vahala}}},\,\href {\doibase 10.1364/OE.22.014559} {\bibfield
  {journal} {\bibinfo  {journal} {Opt. Express}\ }\textbf {\bibinfo {volume}
  {22}},\ \bibinfo {pages} {14559} (\bibinfo {year} {2014})}\BibitemShut
  {NoStop}
\bibitem [{\citenamefont {Qiu}\ \emph {et~al.}(2019)\citenamefont
  {Shomroni},~\citenamefont {Ioannou},~\citenamefont {Piro},~\citenamefont {Malz},~\citenamefont {Nunnenkamp},\ and\~\citenamefont
  {Kippenberg}}]{Qiu2019}%
  \BibitemOpen
  \bibfield  {author} {\bibinfo {author} {\bibfnamefont {L.}~\bibnamefont
  {Qiu}}, \bibinfo {author} {\bibfnamefont {I.}~\bibnamefont
  {Shomroni}}, \bibinfo {author} {\bibfnamefont {M.~A.}~\bibnamefont
  {Ioannou}}, \bibinfo {author} {\bibfnamefont {N.}~\bibnamefont
  {Piro}}, \bibinfo {author} {\bibfnamefont {D.}~\bibnamefont
  {Malz}}, \bibinfo {author} {\bibfnamefont {A.}~\bibnamefont
  {Nunnenkamp}},\ and\ \bibinfo {author} {\bibfnamefont {T.~J.}~\bibnamefont
  {Kippenberg}},\ }\href {\doibase 10.1103/PhysRevA.100.053852} {\bibfield
  {journal} {\bibinfo  {journal} {Phys. Rev. A}\ }\textbf {\bibinfo {volume}
  {100}},\ \bibinfo {pages} {053852} (\bibinfo {year} {2019})}\BibitemShut
  {NoStop}%
\bibitem [{\citenamefont {Ma}\ \emph {et~al.}(2021)\citenamefont
  {Guccione},~\citenamefont {Lecamwasa},~\citenamefont {Qin},~\citenamefont {Campbell},~\citenamefont {Buchler},\ and\~\citenamefont
  {Lam}}]{Ma2021}%
  \BibitemOpen
  \bibfield  {author} {\bibinfo {author} {\bibfnamefont {J.}~\bibnamefont
  {Ma}}, \bibinfo {author} {\bibfnamefont {G.}~\bibnamefont
  {Guccione}}, \bibinfo {author} {\bibfnamefont {R.}~\bibnamefont
  {Lecamwasam}}, \bibinfo {author} {\bibfnamefont {J.}~\bibnamefont
  {Qin}}, \bibinfo {author} {\bibfnamefont {G.~T.}~\bibnamefont
  {Campbell}}, \bibinfo {author} {\bibfnamefont {B.~C.}~\bibnamefont
  {Campbell}},\ and\ \bibinfo {author} {\bibfnamefont {P.~K.}~\bibnamefont
  {Lam}},\ }\href {\doibase 10.1364/OPTICA.412182} {\bibfield
  {journal} {\bibinfo  {journal} {OPTICA}\ }\textbf {\bibinfo {volume}
  {8}},\ \bibinfo {pages} {177} (\bibinfo {year} {2021})}\BibitemShut
  {NoStop}%
\bibitem [{\citenamefont {Allain}\ \emph {et~al.}(2021)\citenamefont
  {Guha}, \citenamefont{Baker}, \citenamefont{Parrain}, \citenamefont{Lema\^itre}, \citenamefont{Leo}, \ and\~\citenamefont
  {Favero}}]{Allain2021}%
  \BibitemOpen
  \bibfield  {author} {\bibinfo {author} {\bibfnamefont {P.~E.}\ \bibnamefont
  {Allain}}, \bibinfo {author} {\bibfnamefont {B.}\ \bibnamefont
  {Guha}}, \bibinfo {author} {\bibfnamefont {C.}\ \bibnamefont
  {Baker}}, \bibinfo {author} {\bibfnamefont {D.}\ \bibnamefont
  {Parrain}}, \bibinfo {author} {\bibfnamefont {A.}\ \bibnamefont
  {Lema\^{i}tre}}, \bibinfo {author} {\bibfnamefont {G.}\ \bibnamefont
  {Leo}},\ and\ \bibinfo {author} {\bibfnamefont {I.}\ \bibnamefont
  {Favero}},\ }\href {\doibase 10.1103/PhysRevLett.126.243901} {\bibfield
  {journal} {\bibinfo  {journal} {Phys. Rev. Lett.}\ }\textbf {\bibinfo {volume}
  {126}},\ \bibinfo {pages} {243901} (\bibinfo {year} {2021})}\BibitemShut
  {NoStop}%
\bibitem [{\citenamefont {Hassett}(2007)}]{Hassett2007}%
  \BibitemOpen
  \bibinfo {author} {\bibfnamefont {B.}~\bibnamefont {Hassett}},\ \href {\doibase 10.1017/CBO9780511755224} { \emph {\bibinfo {title} {Introduction into Algebraic Geometry}}}\ (\bibinfo  {publisher} {Cambridge University Press, Cambridge},\
  \bibinfo {year} {2007})\BibitemShut {NoStop}%
\bibitem [{\citenamefont {Li}\ \emph {et~al.}(2010)\citenamefont {Wang}, \citenamefont {Su},   \citenamefont{Yan},\ and\~\citenamefont {Qiu}}]{Li2010}%
  \BibitemOpen
  \bibfield  {author} {\bibinfo {author} {\bibfnamefont {Q.}~\bibnamefont
  {Li}}, \bibinfo {author} {\bibfnamefont {T.}~\bibnamefont
  {Wang}}, \bibinfo {author} {\bibfnamefont {Y.}~\bibnamefont
  {Su}}, \bibinfo {author} {\bibfnamefont {M.}~\bibnamefont
  {Yan}},\ and\ \bibinfo {author} {\bibfnamefont {M.}~\bibnamefont
  {Qiu}}},\,\href {\doibase 10.1364/OE.18.008367} {\bibfield
  {journal} {\bibinfo  {journal} {Opt. Express}\ }\textbf {\bibinfo {volume}
  {18}},\ \bibinfo {pages} {8367} (\bibinfo {year} {2010})}\BibitemShut
  {NoStop}
\bibitem [{\citenamefont {Tsvirkun}\ \emph {et~al.}(2015)\citenamefont {Surrente}, \citenamefont {Raineri}, \citenamefont{Beaudoin}, \citenamefont{Raj}, \citenamefont{Sagnes}, \citenamefont{Robert-Philip},\ and\~\citenamefont {Braive}}]{Tsvirkun2015}%
  \BibitemOpen
  \bibfield  {author} {\bibinfo {author} {\bibfnamefont {V.}~\bibnamefont
  {Tsvirkun}}, \bibinfo {author} {\bibfnamefont {A.}~\bibnamefont
  {Surrente}}, \bibinfo {author} {\bibfnamefont {F.}~\bibnamefont
  {Raineri}}, \bibinfo {author} {\bibfnamefont {G.}~\bibnamefont
  {Beaudoin}}, \bibinfo {author} {\bibfnamefont {R.}~\bibnamefont
  {Raj}}, \bibinfo {author} {\bibfnamefont {I.}~\bibnamefont
  {Sagnes}}, \bibinfo {author} {\bibfnamefont {I.}~\bibnamefont
  {Robert-Philip}},\ and\ \bibinfo {author} {\bibfnamefont {R.}~\bibnamefont
  {Braive}}},\,\href {\doibase 10.1038/srep16526} {\bibfield
  {journal} {\bibinfo  {journal} {Sci. Rep}\ }\textbf {\bibinfo {volume}
  {5}},\ \bibinfo {pages} {16526} (\bibinfo {year} {2015})}\BibitemShut
  {NoStop}
\bibitem [{\citenamefont {Brunstein}\ \emph {et~al.}(2009)\citenamefont {Braive}, \citenamefont {Hostein}, \citenamefont{Beveratos}, \citenamefont{Robert-Philip}, \citenamefont{Sagnes}, \citenamefont{Karle}, \citenamefont{Yacomotti}, \citenamefont{Levenson}, \citenamefont{Moreau}, \citenamefont{Tessier},\ and\~\citenamefont {De Wilde}}]{Brunstein2009}%
  \BibitemOpen
  \bibfield  {author} {\bibinfo {author} {\bibfnamefont {M.}~\bibnamefont
  {Brunstein}}, \bibinfo {author} {\bibfnamefont {R.}~\bibnamefont
  {Braive}}, \bibinfo {author} {\bibfnamefont {R.}~\bibnamefont
  {Hostein}}, \bibinfo {author} {\bibfnamefont {A.}~\bibnamefont
  {Beveratos}}, \bibinfo {author} {\bibfnamefont {I.}~\bibnamefont
  {Robert-Philip}}, \bibinfo {author} {\bibfnamefont {I.}~\bibnamefont
  {Sagnes}}, \bibinfo {author} {\bibfnamefont {T.~J.}~\bibnamefont
  {Karle}}, \bibinfo {author} {\bibfnamefont {A.~M.}~\bibnamefont
  {Yacomotti}}, \bibinfo {author} {\bibfnamefont {J.~A.}~\bibnamefont
  {Levenson}}, \bibinfo {author} {\bibfnamefont {V.}~\bibnamefont
  {Moreau}}, \bibinfo {author} {\bibfnamefont {G.}~\bibnamefont
  {Tessier}},\ and\ \bibinfo {author} {\bibfnamefont {Y.}~\bibnamefont
  {De Wilde}}},\,\href {\doibase 10.1364/OE.17.017118} {\bibfield
  {journal} {\bibinfo  {journal} {Opt. Express}\ }\textbf {\bibinfo {volume}
  {17}},\ \bibinfo {pages} {17118} (\bibinfo {year} {2009})}\BibitemShut
  {NoStop}
\bibitem [{\citenamefont {Genes}\ \emph {et~al.}(2009)\citenamefont
  {Mari},\ \citenamefont
  {Vitali},\ and\~\citenamefont {Tombesi}}]{Genes2009}%
  \BibitemOpen
  \bibfield  {author} {\bibinfo {author} {\bibfnamefont {C.}~\bibnamefont
  {Genes}}, \bibinfo {author} {\bibfnamefont {A.}~\bibnamefont
  {Mari}}, \bibinfo {author} {\bibfnamefont {D.}~\bibnamefont
  {Vitali}}},\ and\ \bibinfo {author} {\bibfnamefont {P.}~\bibnamefont
  {Tombesi}},\,\href {\doibase 10.1016/S1049-250X(09)57002-4} {\bibfield
  {journal} {\bibinfo  {journal} {Adv. At. Mol. Opt. Phys.}\ }\textbf {\bibinfo {volume}
  {57}},\ \bibinfo {pages} {33} (\bibinfo {year} {2009})}\BibitemShut
  {NoStop}
\bibitem [{\citenamefont {Karuza}\ \emph {et~al.}(2012)\citenamefont
  {Molinelli},\ \citenamefont {Galassi},\ \citenamefont {Biancofiore},\ \citenamefont
  {Natali},\ \citenamefont {Tombesi},\ \citenamefont {Di Giuseppe},\ 
  and\~\citenamefont {Vitali}}]{Karuza2012}%
  \BibitemOpen
  \bibfield  {author} {\bibinfo {author} {\bibfnamefont {M.}~\bibnamefont
  {Karuza}}, \bibinfo {author} {\bibfnamefont {C.}~\bibnamefont
  {Molinelli}}, \bibinfo {author} {\bibfnamefont {M.}~\bibnamefont
  {Galassi}}, \bibinfo {author} {\bibfnamefont {C.}~\bibnamefont
  {Biancofiore}}, \bibinfo {author} {\bibfnamefont {P.}~\bibnamefont
  {Natali}}, \bibinfo {author} {\bibfnamefont {A.}~\bibnamefont
  {Tombesi}}, \bibinfo {author} {\bibfnamefont {G.}~\bibnamefont
  {Di Giuseppe}}},\ and\ \bibinfo {author} {\bibfnamefont {D.}~\bibnamefont
  {Vitali}},\,\href {\doibase 10.1088/1367-2630/14/9/095015} {\bibfield
  {journal} {\bibinfo  {journal} {New. J. Phys.}\ }\textbf {\bibinfo {volume}
  {14}},\ \bibinfo {pages} {095015} (\bibinfo {year} {2012})}\BibitemShut
  {NoStop}
\bibitem [{\citenamefont {Kloeden}(1992)}]{Kloeden1992}%
  \BibitemOpen
  \bibinfo {author} {\bibfnamefont {P.~E.}~\bibnamefont {Kloeden}}\ and\ \bibinfo {author} {\bibfnamefont {E.}~\bibnamefont
  {Platen}},\ \href {\doibase 10.1007/978-3-662-12616-5} { \emph {\bibinfo {title} {Numerical Solution of Stochastic Differential Equations}}}\ (\bibinfo  {publisher} {Springer-Verlag, Berlin-Heidelberg},\
  \bibinfo {year} {1992})\BibitemShut {NoStop}%
\bibitem [{\citenamefont {Pelc}\ \emph {et~al.}(2014)\citenamefont
  {Rivoire},\ and\~\citenamefont{Vo},\ and\~\citenamefont{Santori},
  \ and\~\citenamefont{Beausoleil},\ and\~\citenamefont{Fattal}}]{Pelc:14}%
  \BibitemOpen
  \bibfield  {author} {\bibinfo {author} {\bibfnamefont {J.~S.}~\bibnamefont
  {Pelc}}, \bibinfo {author} {\bibfnamefont {K.}\ \bibnamefont
  {Rivoire}},\ and\ \bibinfo {author} {\bibfnamefont {S.}~\bibnamefont
  {Vo}},\ and\ \bibinfo {author} {\bibfnamefont {C.}~\bibnamefont
  {Santori}},\ and\ \bibinfo {author} {\bibfnamefont {D.~A.}~\bibnamefont
  {Fattal}},\ and\ \bibinfo {author} {\bibfnamefont {R.~G.}~\bibnamefont
  {Beausoleil}},\ }\href {\doibase 10.1364/OE.22.003797} {\bibfield
  {journal} {\bibinfo  {journal} {Opt. Express}\ }\textbf {\bibinfo {volume}
  {22}},\ \bibinfo {pages} {3797--3810} (\bibinfo {year} {2014})}\BibitemShut
  {NoStop}%
\end{thebibliography}
\end{document}